\documentclass[%
 amsmath,amssymb,
 aps,
pre,
]{revtex4-1}

\usepackage{graphicx}
\usepackage{dcolumn}
\usepackage{bm}
\usepackage{color,soul}
\usepackage{subfig}
\usepackage{xcolor}
\usepackage{bbm}

\newlength{\mywidth}
\begin{document}

\title{Breaking of Josephson junction oscillations and onset of quantum turbulence in Bose--Einstein condensates}

\author{Adam Griffin}
\affiliation{Mathematics Institute, The University of Warwick, Coventry, CV4 7AL, United Kingdom}
\email{a.griffin.1@warwick.ac.uk}
\author{Sergey Nazarenko}
\affiliation{Institut de Physique de Nice, Universit\'e C\^ote d'Azur, CNRS, Nice, France}%
\author{Davide Proment}%
\affiliation{School of Mathematics, University of East Anglia, Norwich Research Park, Norwich, NR4 7TJ, United Kingdom}%

\date{\today}

\begin{abstract}
We analyse the formation and the dynamics of quantum turbulence in a two-dimensional Bose--Einstein condensate with a Josephson junction barrier modelled using the Gross--Pitaevskii equation. 
We show that a sufficiently high initial superfluid density imbalance leads to randomisation of the dynamics and generation of turbulence, namely, the formation of a quasi-1D dispersive shock consisting of a train of grey solitons that eventually breakup into chains of distinct quantised vortices of alternating vorticity followed by random turbulent flow. 
The Josephson junction barrier allows us to create two turbulent regimes: acoustic turbulence on one side and vortex turbulence on the other.
Throughout the dynamics, a key mechanism for mixing these two regimes is the transmission of vortex dipoles through the barrier: we analyse this scattering process in terms of the barrier parameters, sound emission and vortex annihilation. 
Finally, we discuss how the vortex turbulence evolves for long times, presenting the optimal configurations for the density imbalance and barrier height in order to create the desired turbulent regimes which last as long as possible. 

\end{abstract}

\maketitle


\section{Introduction}\label{INTRO}


The Josephson junction (JJ) is an experimental set-up designed to showcase the Josephson effect \cite{JOSEPHSON1962251}. 
This quantum mechanical effect, which describes particle tunnelling through a barrier and periodic oscillations, is well studied in the context of Bose--Einstein condensates (BECs) in both theory \cite{refId0, PhysRevA.95.023627, PitStrBEC2003} and experiments \cite{Cataliotti843, PhysRevLett.95.010402, Levy:2007yu, Valtolina1505, PhysRevLett.118.230403}. 
In this article, we discuss the dynamics of a BEC JJ which is pushed to its limit and ``goes bad''. Namely, we study the regime where the periodic oscillations in the superfluid density break down and quantum turbulence arises in the system. 

 Current methods to create vortex turbulence include optical spoons \cite{PhysRevLett.84.806} and shaking confining traps \cite{Navon:2016aa}. These methods require much energy to create a single vortex in fluids with high density. Moreover, if the density is low the resulting vortices annihilate very quickly; therefore, the creation rate will be similar to the annihilation rate. We propose a method to create vortex dipoles with initial imbalance and sustain them by using a Kibble-Zurek \cite{Kibble:1980aa,Zurek:1985aa} like mechanism to prolong the vortex turbulence, that is, create vortices in a region of low density and then increase the density in a controlled manner to maintain the topological defects and decrease the relative strength of the acoustic waves.

The Josephson junction consitsts of a barrier or weak link separating two wells of superfluid or superconductor. We consider two wells separated by a potential barrier with an initial density imbalance between the wells. As the system evolves the fluid moves through or over the barrier, which leads to oscillatory dynamics. We show that when the initial density imbalance is pushed to high values, the regime of regular oscillations breaks down. The system then exibits chaotic behaviour with interesting nonlinear dynamics, consisting of chaotic motion of vortices coupled with turbulent acoustic waves, following the break down of a soliton train caused by a dispersive shock \cite{Mossman:2018aa}. Studies on critical parameters for vortex generation have been undertaken \cite{xhani2019critical}. However, these studies do not consider high numbers of vortices which is the case in turbulence.


In place of the predictions by Josephson, we see interesting turbulence characterised by a separation of acoustic and vortex turbulence. Such chaotic dynamics appear when an initial train of solitons is formed then breaks down, causing the generation of vortices which is due to the instability of quasi-1D solitons. 

We demonstrate that the Josephson junction is a nonlinear system and that by tuning experimental parameters, we can produce rich and controllable non-linear behaviour, which gives an ideal set up for turbulence. Readily available experimental apparata in BECs \cite{Cataliotti843,Navon:2016aa}, atomic vapors \cite{PhysRevLett.120.055301} and photorefractive crystals \cite{Bellec} can implement such a system. In similar experimental systems the emergence of a few vortices has recently been witnessed \cite{Burchianti:2018aa}, our study differs as we foccus on finding specific parameters that produce the largest number of vortices which are sustained for the longest time so that we can observe isotropic vortex turbulence. We show that a crucial element in the turbulent dynamics in such a system is due to the interaction of vortex dipoles with the barrier. These interactions are responsible for the separation of turbulence within the two wells, with one retaining most the vortices and the other containing weak acoustic wave turbulence \cite{Nazarenko_2011}. Certain parameters (for instance the incidence angle) control the ability for vortices to cross the barrier \cite{Mironov2013,0953-4075-47-13-135301}. \\



\section{The mathematical model}\label{TMM}

To model the JJ theoretically we perform direct numerical simulations (DNS) of the Gross--Pitaevskii (GP) equation. The GP equation describes the dynamics of a BEC made of a dilute ultra-cold gas of bosons \cite{Gross1961,Pit1961}. 
For simplicity we consider the case of a quasi-two dimensional BEC, that is, we simulate the GP equation in two spatial dimensions as well as time. Lengths are expressed in units of the healing length $\xi = \hbar/\sqrt{2m\rho_0 g_{2D}}$ where $\rho_0 $ is the mean density, $ g_{2D} = \frac{\sqrt{8\pi}\hbar^2 a_s}{m a_z}$ is the effective two-dimensional interaction constant between bosons of mass $m$, $\mu$ is the chemical potential of the system, $a_s$ is the s-wave scattering length of the particle interactions and $a_z$ is a length scale corresponding to the confinement to 2D.
Time is rescaled by $\sqrt{2}\xi/c$, where $ c=\sqrt{\rho_0g_{2D}/m} $ is the speed of large scale density/phase fluctuations (sound) in the bulk. The external potential $V$ is given in units of $\rho_0g_{2D}$. 

The non-dimensional form of the GP equation reads as follows:
\begin{equation}
i \frac{\partial\psi }{\partial t}=\left( -\nabla^{2} + V({\bf x},t) +\left|\psi({\bm x},t) \right|^{2}  \right ) \psi({\bm x},t) \, .
\label{eqn:gpe}
\end{equation}
The complex wave function $\psi(x,y,t)$ is the BEC order parameter and can be expressed in fluid like variables via the Madelung transformation $\psi =\sqrt{ \rho({\bf x},t)}e^{i\phi({\bf x},t)}$, where $\rho({\bf x},t) = |\psi({\bf x},t)|^2$ and ${\bf v}({\bf x},t) = 2\nabla \phi({\bf x},t)$ are the density and velocity of the superfluid respectively. In our dimensionless variables we rescale density by the initial density $\rho_0$.

Our aim is to model an elongated JJ domain. We require that the domain is large with respect to the healing length in order to observe the formation of several quantised vortices and, eventually, fully developed quantum turbulence.
We thus choose a two-dimensional spatial domain ${\bf x}=(x,y)$, with $ x =  [- 256\xi , 256\xi] $ and $ y =  [-128\xi ,128\xi] $, setting a computational uniform grid of spacing $0.25\xi$. We have Dirichlet boundary conditions with $\psi=0$ at the boundary, this is effectively confining the fluid in an abrupt rectangular trap with an external potential at the boundary with infinite strength.
The JJ barrier is modelled using an external potential $V_{JJ}({\bf x},t)$ that separates the domain into two equally sized boxes labelled $B_L$ and $B_R$, corresponding to the position left or right of the potential. 
The potential is given by a Gaussian function centred at $x=0$ and stretched along the entire y-axis. 
To create the initial superfluid density imbalance needed to trigger the JJ oscillations, an extra non-zero external potential (almost) uniform $ V_d $ is present in the right box whilst we minimise the energy of the system.
Mathematically, the JJ barrier thus results in


\begin{equation}
V_{JJ}(x,y) =
\begin{cases}
V_{0}e^{-\frac{x^2}{\sigma^2}}+V_{d} \tanh(x), \hspace{4mm} \text{To create initial conditions,}\\
V_{0}e^{-\frac{x^2}{\sigma^2}}, \hspace{25mm} \text{For the dynamics,} \\
\end{cases}\label{POTi}
\end{equation}

where $ V_0 $ and $ \sigma $ control the intensity and the width of the JJ barrier respectively and $V_d$ sets the initial density imbalance.
In all the simulations reported in this work we keep $\sigma = 1.2 $ so that the width of the barrier is always similar to the healing length of the system.

To create the initial condition, we use imaginary time propagation (ITP). This method involves evolving the GP model with the substitution $t=-i \tau$ while imposing conservation of the number of particles to effectively minimise the energy of the system. 
Once the desired energy stagnation has been reached, we call the obtained state the {\it ground state}, and evolve the system in real-time with the ground state as the initial conditions. Because we also alter $ V_{JJ} $, the initial condition is no longer the ground state of the system that we then propagate; therefore, dynamics follow. 

The GP equation \eqref{eqn:gpe} is integrated using the 4th order central finite difference method in space and a Runge-Kutta 4th order method for the time-stepping scheme. 
Due to the nature of the method, only waves with wavelength of the order of half the healing length or higher are well resolved; smaller waves will be numerically dissipated. 
This natural dissipation of high-energy, high-frequency waves is ideal for such a study as dissipation occurs in experiments similarly due to the interaction with the thermal cloud and/or bosons losses due to finite-amplitude confining potential \cite{PhysRevA.57.4057,PhysRevLett.119.190404}. This however, is not controlled dissipation and is set by the choice of grid size. To improve the model one could include dissipation in the numerics by using a de-aliased spectral scheme along with hyper-viscosity; however, such schemes are periodic by nature and may introduce unwanted Gibbs phenomenon due to the hard boundaries which simulate the trapping potential and undesirable periodicity effects. 

\section{Results\label{sec:results}}

\subsection{Measurable quantaties}\label{NR}
We define the relative superfluid density imbalance
\begin{equation}
Z(t)= \frac{N_L(t)-N_R(t)}{N_L(t) + N_R(t)}
\end{equation}
using the total number of particles
\begin{equation}
N_i= \int_{B_i} |\psi(x,y,t)|^2 dxdy 
\end{equation}
per box $B_i$ where $i$ is an index for the box left or right of the separating potential,
Note that the total number of particles $N=N_L+N_R$ is an integral of motion and it is numerically conserved in all simulations up to 0.002\%. We also define the initial density imbalance $Z_0 = Z(t=0)$.
Analogously, the energy per box reads
\begin{equation}
E_i(t) = \int_{B_i} |\nabla\psi(x,y,t)|^2 +V(x,y)|\psi(x,y,t)|^2+\frac{1}{2}|\psi(x,y,t)|^4  dxdy \,.\label{EN}
\end{equation}
The total energy $E = E_L + E_R $ is also an integral of motion, but due to the intrinsic high-frequency numerical dissipation, its value decreases in time as much as 28\% when $Z_0=0.49$ around 10\% when $Z_0=0.88$. For the results in section \ref{VDSWB}, where there is no acoustic turbulence, the energy is conserved to $0.0015\%$. The energy is naturally decomposed in \eqref{EN}, with the second term corresponding to energy from the external potential and the third term the internal energy of the fluid. We can further decompose the first term into kinetic and quantum energy by applying the Madelung transformation $\psi=\sqrt{\rho}e^{i\phi}$ to the first term

\begin{align}
|\nabla \psi| ^2 = |\nabla \phi |^2\rho + |\nabla \sqrt{\rho}|^2,
\end{align}
where the first term corresponds to the kinetic energy density and the second term is the so-called quantum energy density. By performing a Helmholtz decomposition on the kinetic energy density we can further decompose into the compressible and incompressible energies. That is, $\varepsilon_{kin} = \varepsilon_{kin}^{c} + \varepsilon_{kin}^{i}$, where the incompressible component of the energy corresponds to the vector field satisfying $\nabla \cdot (\sqrt{\rho}{ \bf v})^{i} = 0$. Further details on the calculation can be found in \cite{Numasato:2010aa}. Thus, the energy can now be written as follows:

\begin{equation}
E_i(t) = \int_{B_i} \varepsilon^{c}_{kin} +\varepsilon^{i}_{kin}+ |\nabla \sqrt{\rho}|^2+V(x,y)|\psi(x,y,t)|^2+\frac{1}{2}|\psi(x,y,t)|^4  dxdy \,.\label{EN2}
\end{equation}

The total incompressible energy $E^{in}_{kin} = \int_{B_L +B_R} \varepsilon_{kin}^{in} dxdy$ is a measure of the energy in large-scale incompressible potential flow and vortices, whereas the total compressible energy $E^{c}_{kin}= \int_{B_L +B_R} \varepsilon_{kin}^{c} dxdy$ is the energy in the acoustic component.


It is also convenient to define the {\emph{local}} healing length for each box as 
\begin{equation}
\xi_i (t) = \sqrt{\frac{L^2}{N_i(t)}}.
\end{equation}
We will call the natural healing length of the system $\xi$, that is, the healing length if the initial density imbalance is set to zero. The length of each box in units of $\xi$ is given by $L$.

Finally, it is instructive to measure the total number of vortices 
\begin{equation}
N_V(t)=N_{V_L}(t) +N_{V_R}(t)
\end{equation}
in the system versus time, with $N_{V_L}$ and $N_{V_R}$ the number for the left and right boxes respectively.
Each quantised vortex is numerically identified using the pseudo-vorticity defined as follows,
\begin{align}
\omega_{\rho s} & = \frac{1}{2}\nabla \times \mathbf{j},
\end{align}
with
\begin{align}
\mathbf{j} &=\rho \mathbf{v} =-\frac{i}{2}(\psi^* \nabla \psi - \psi\nabla \psi^*)\\
\end{align}
where $\mathbf{j}$ is the density flux.
Then we find the maxima of $\omega_{\rho s}$ in simply-connected regions ignoring the field below a chosen cut-off value, see \cite{Villois_2016} for further details. 
Ghost vortices are phase fluctuations in large regions where $\psi$ is close to zero, these do not show the same dynamics as hydrodynamical vortices. We add an extra filter to our vortex tracker, namely we only consider vortices with substantial density surrounding them; this is to remove the ghost vortices and to track only the hydrodynamic vortices. The numerical scheme calculates the average density around any point identified by the vortex tracking routine and discards the vortex if the average is below a threshold value.

\subsection{Creation of vortices}\label{CVd}
In the GP model, 1D dark or grey solitons are unstable to transverse perturbations in two spatial dimensions, this instability is known as the snake instability. The instability is a result of the speed of a soliton being proportional to its amplitude and it can be understood by considering the implications of smaller solitons having larger speeds. A small transverse perturbation introduces a local difference in speed of the soliton. Such a difference in speed will cause the soliton to bulge in the direction of motion if the perturbation is negative, or the opposite direction if the perturbation is positive. For instance, if we introduce a bulge in the direction of motion, since the soliton will move perpendicular to its tangent there will be a focusing effect on either side of the perturbation. Since the speed of the soliton is reduced when its amplitude increases, the focused parts of the solitons slow down producing an inverted bulge. The process then continues along the length of the soliton with bulges and inverted bulges forming along the soliton, for more details and mathematical analysis see \cite{Kuznetsov:1988aa}. When a grey-soliton's amplitude becomes as large as the density around it, i.e. points at which $\psi=0$ will appear: phase defects in the form of vortices are nucleated. In the rest of this section we will discuss how we take advantage of this instability to produce vortices. 

\begin{figure}[ht]
\centering
\includegraphics[width=0.45\linewidth]{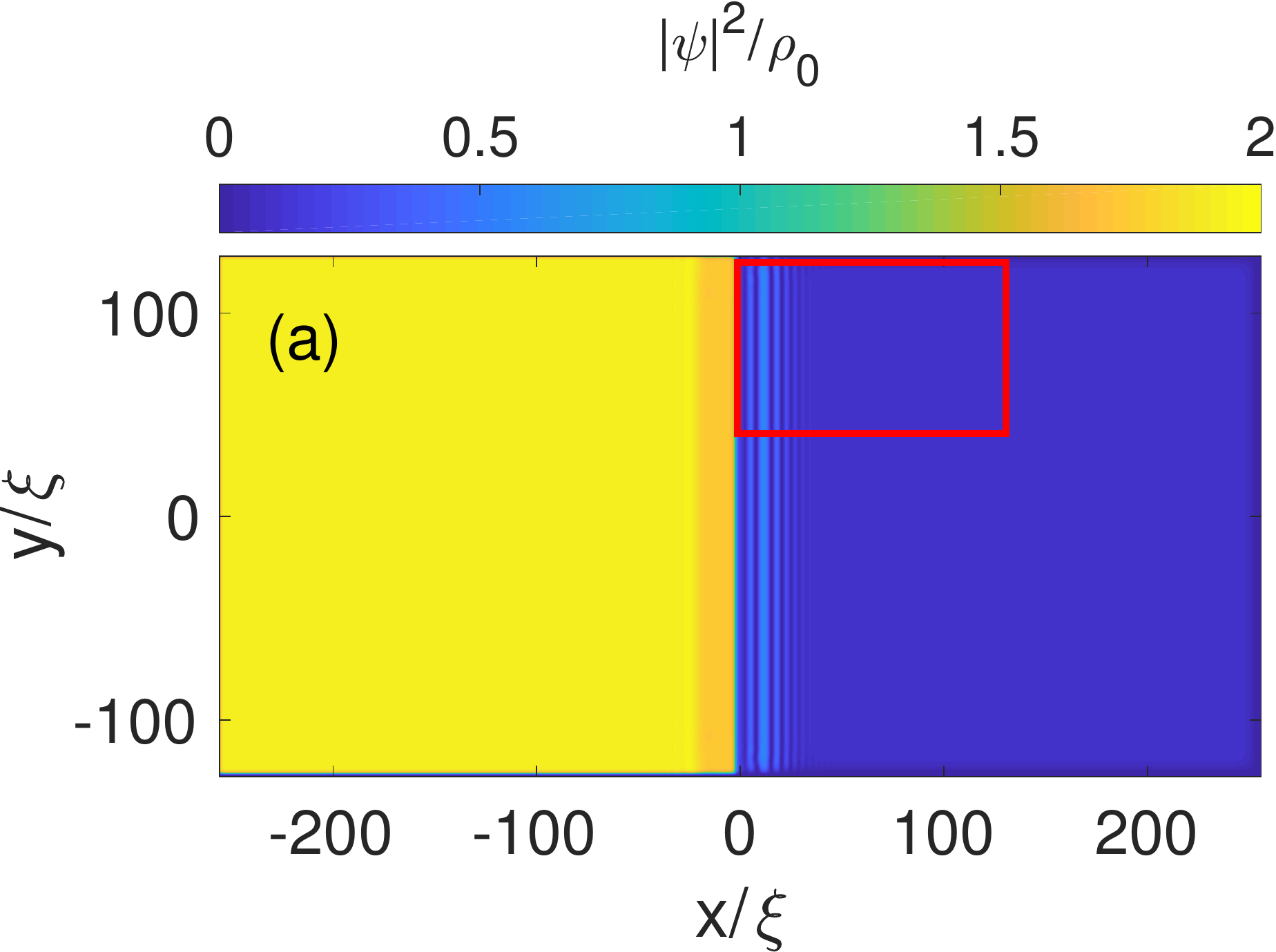}
\includegraphics[width=0.45\linewidth]{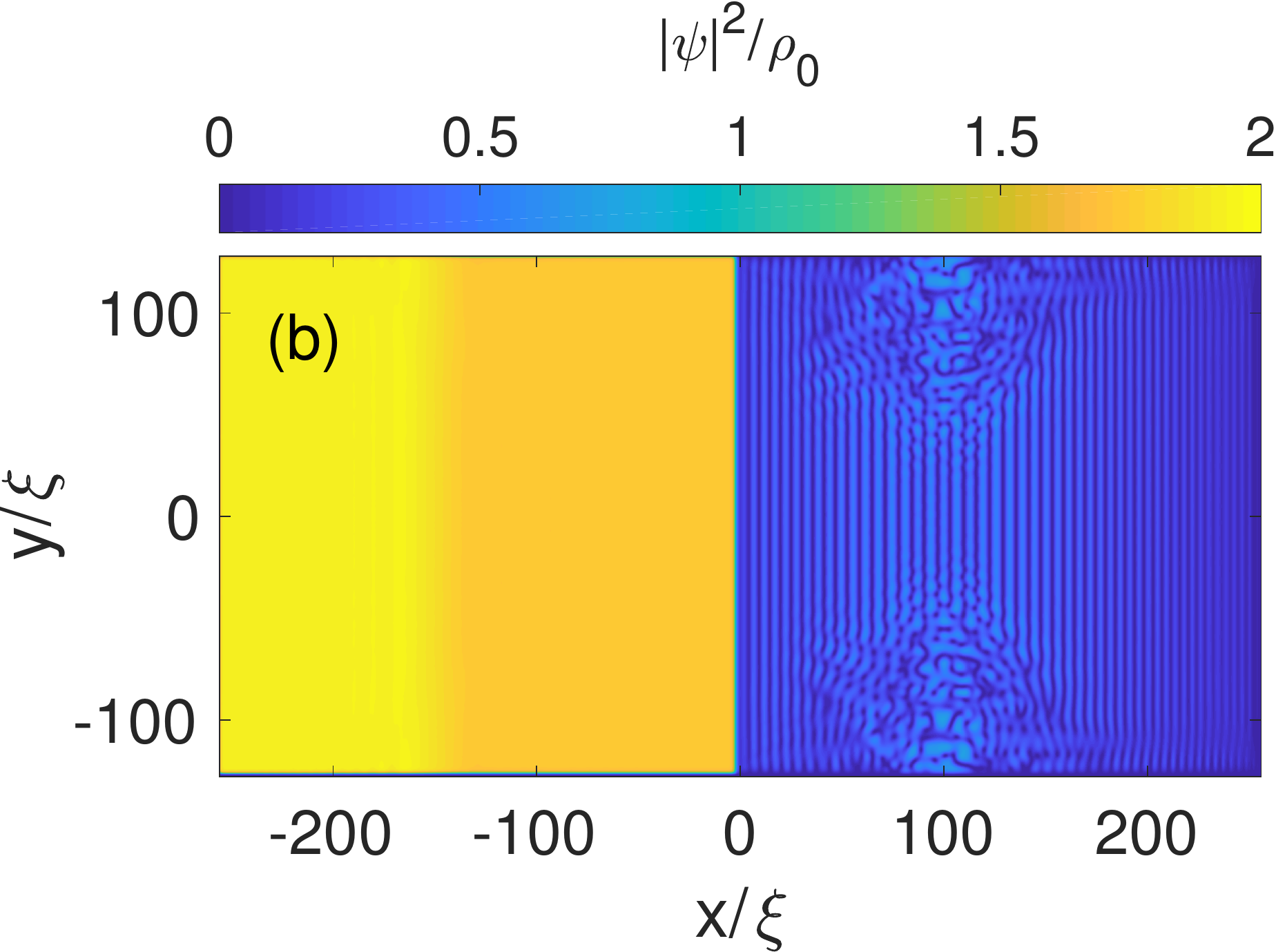}\\
\includegraphics[width=0.45\linewidth]{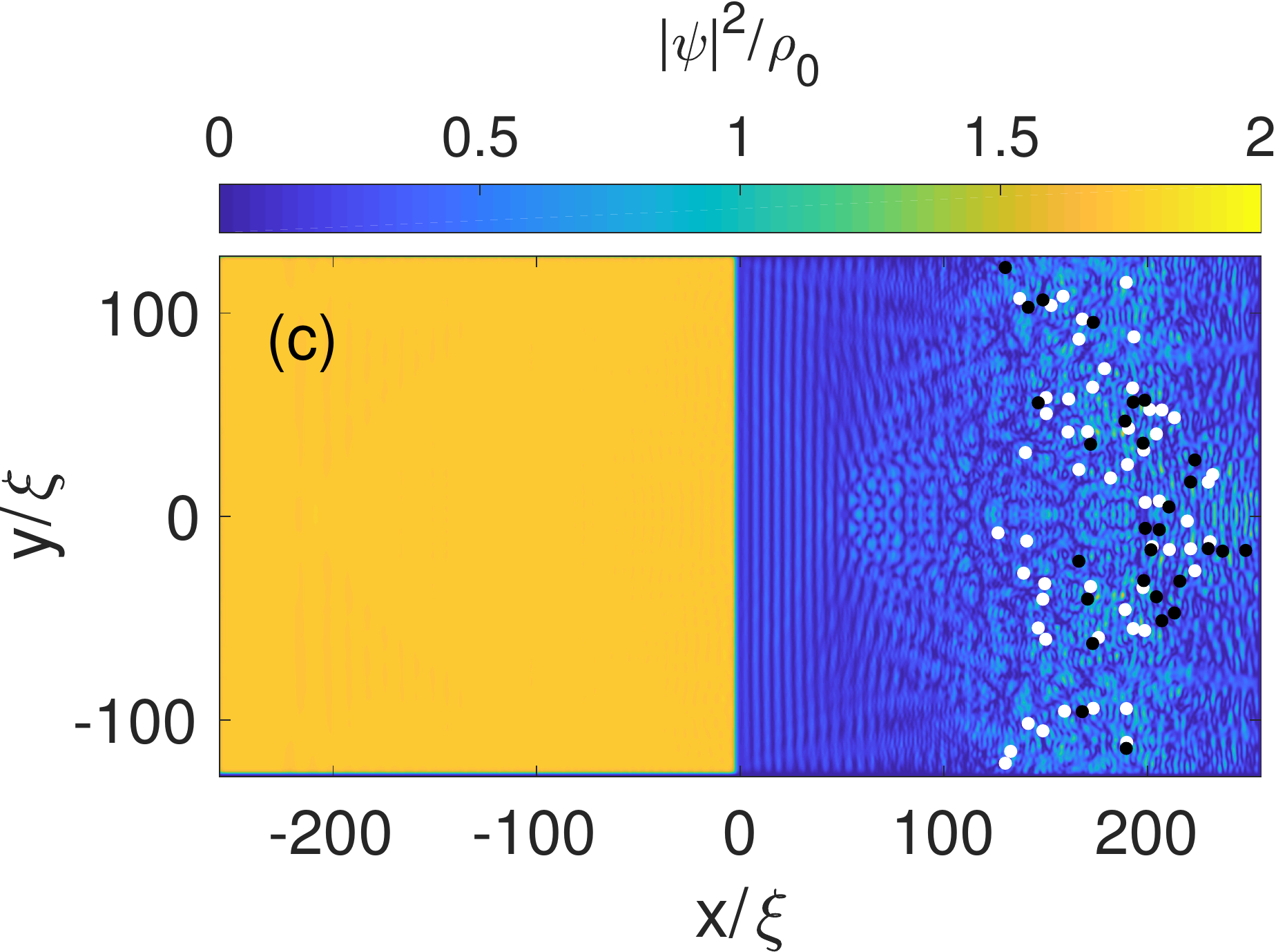}
\includegraphics[width=0.45\linewidth]{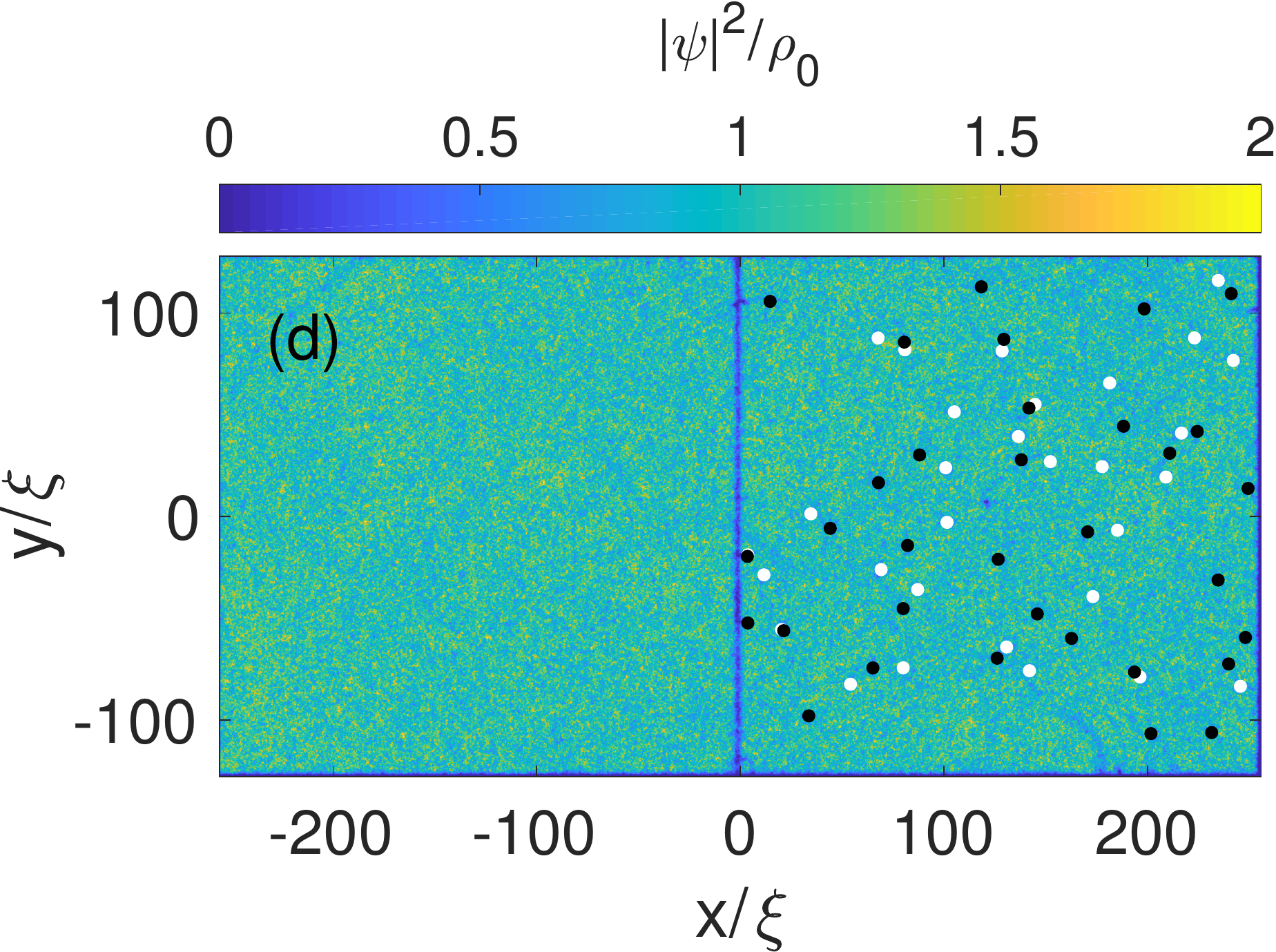}
 \caption{{\it (Color online)} Density fields for simulation with $V_0=1.5/\mu$, $\sigma=1.2/\xi$ and $Z_0 = 0.88$. Panel (a): $t=20 \xi/c$; (b): $t=90\xi/c$, Panel~{(c)}: $t=150\xi/c$, Panel~{(d)}: at $t=3000\xi/c$. The red box in (a) indicates the region which we zoom in on in Fig.~\ref{SOLS}. Positive vortices are shown with white circles and negative with black circles.}\label{DENSL}
 \end{figure}

The initial conditions are chosen such that a train of solitons is produced within a dispersive shock wave in the right well (which has low density $N_L(0)<N_R(0)$), this can be seen in Fig.~\ref{DENSL} which shows an example simulation for the entire domain. Fig.~\ref{DENSL}(a) shows the production of the train of solitons, seen as the stripes in the low density region. The solitons will then decay by the snake instability into alternating signed vortices, the process begins at the boundary and can be seen in Fig.~\ref{DENSL} (b) with later stages in Fig.~\ref{DENSL} (c). Due to the large number of vortices and to the fact that the local density is small, the initial vortices have large cores and do not interact like hydrodynamic vortices. These vortices are often referred to as ghost vortices and are not counted by the tracking algorithm. The continuous flow of solitons carrying density into the right well reduces the local healing length, which in turn transitions the ghost vortices towards hydrodynamic vortices. After the ghost vortices are produced in Fig.~\ref{DENSL} (b), we see chaotic motion with a proliferation of hydrodynamic vortices in Fig.~\ref{DENSL} (c) when the density becomes larger due to the fluid flux from the left box. As the process continues, the healing length tends to $\xi$ in both boxes, the healing length when $N_L=N_R$. Some opposite-signed vortices annihilate which results in continuous decay of the total number of vortices. As a result, the remaining vortices become closer to hydrodynamic vortices because the mean distance between them becomes much greater than $\xi$.  Later stages of the dynamics are shown in Figs.~\ref{DENSL}{(d)} where many of the vortices have decayed.

\begin{figure}[ht]
\centering
\includegraphics[width=0.24\linewidth]{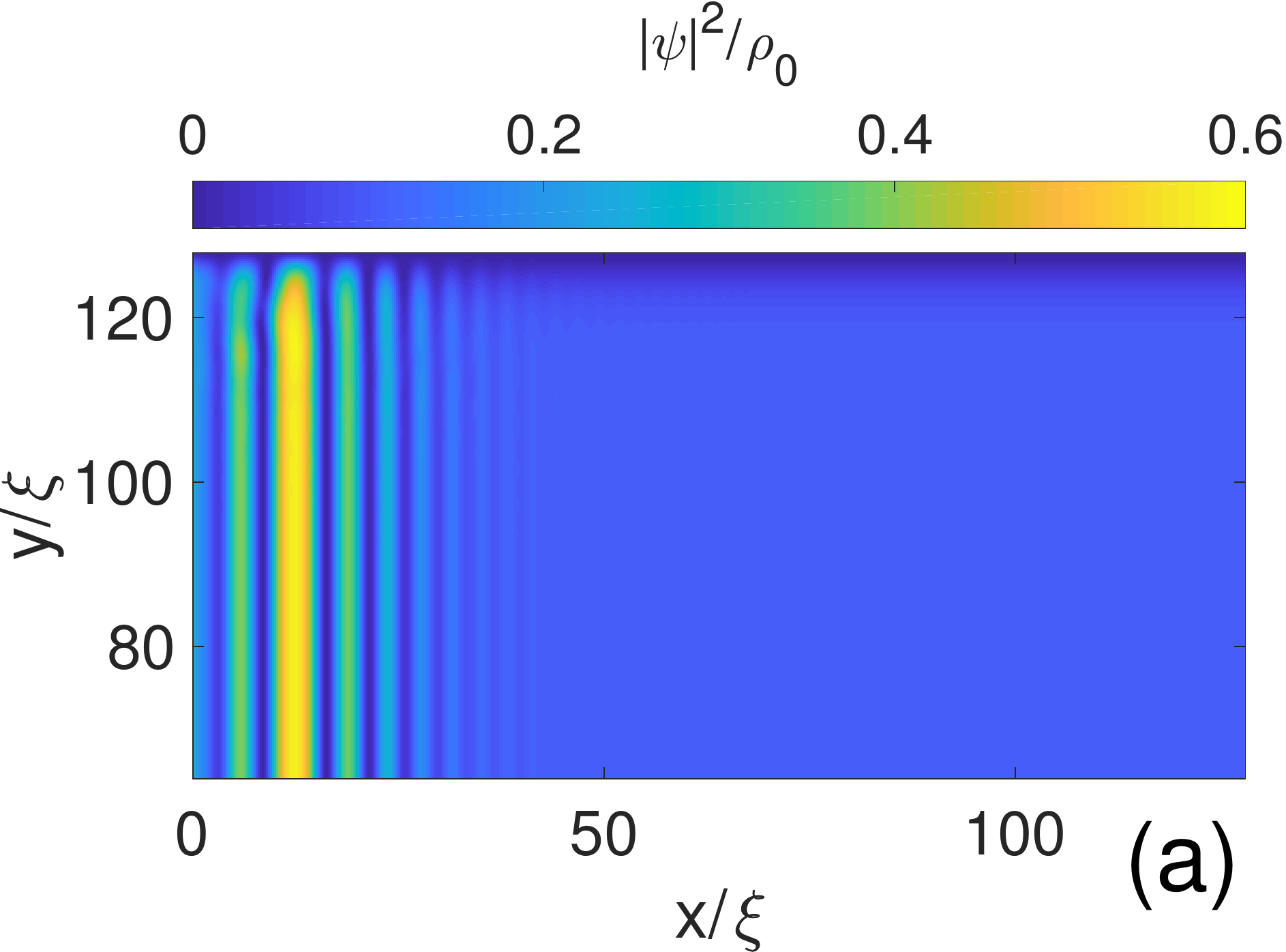}
\includegraphics[width=0.24\linewidth]{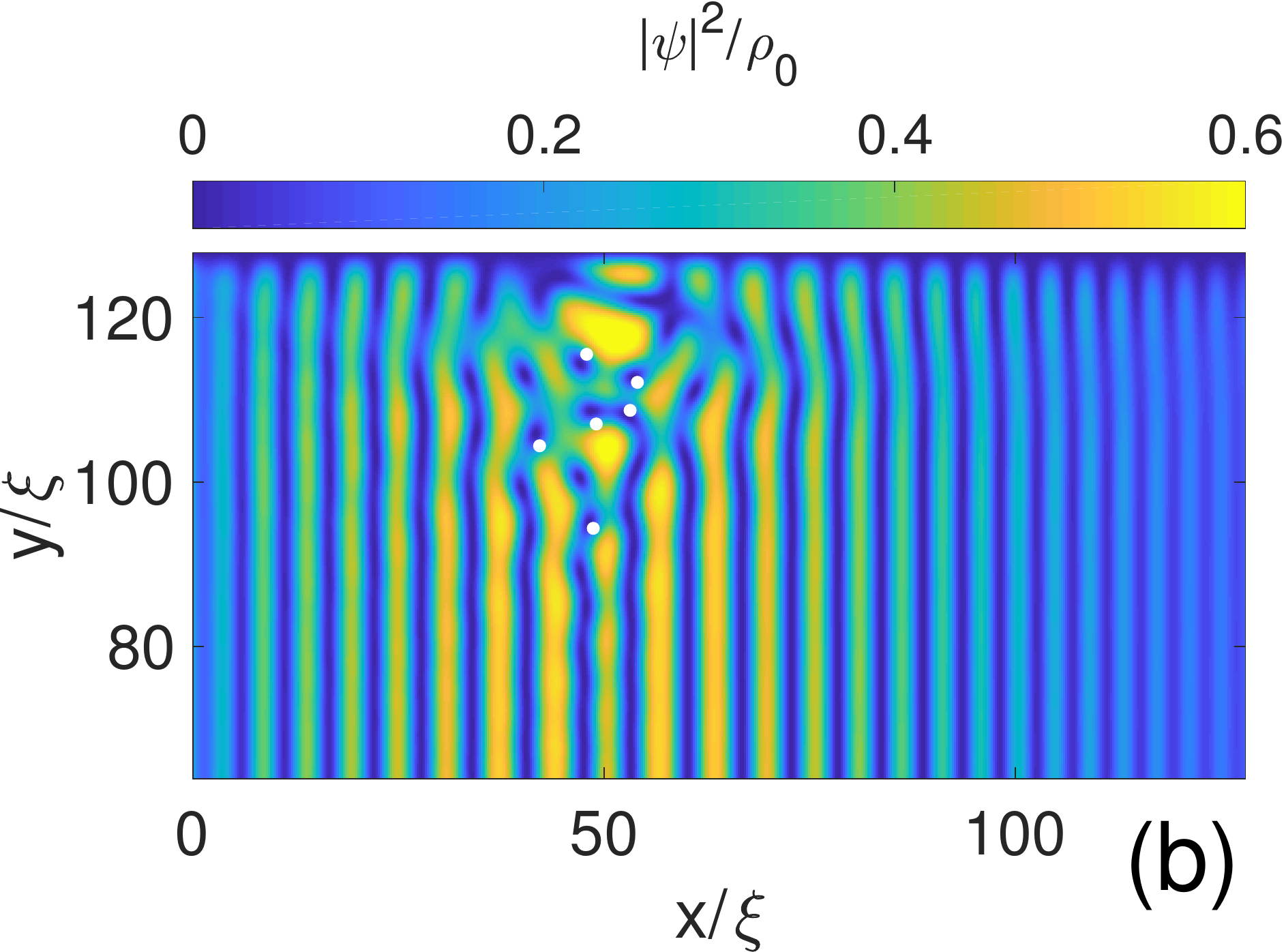}
\includegraphics[width=0.24\linewidth]{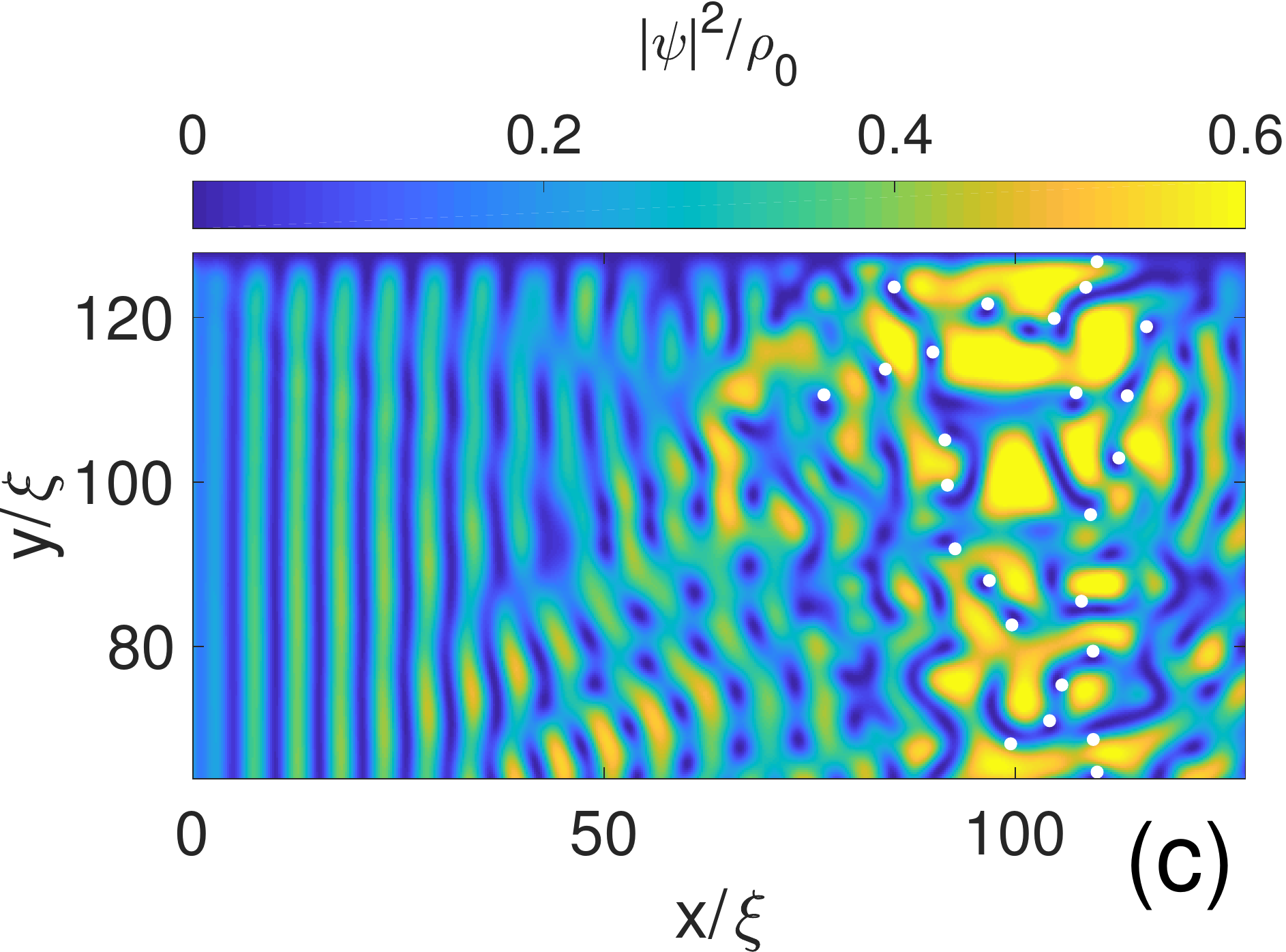}
\includegraphics[width=0.24\linewidth]{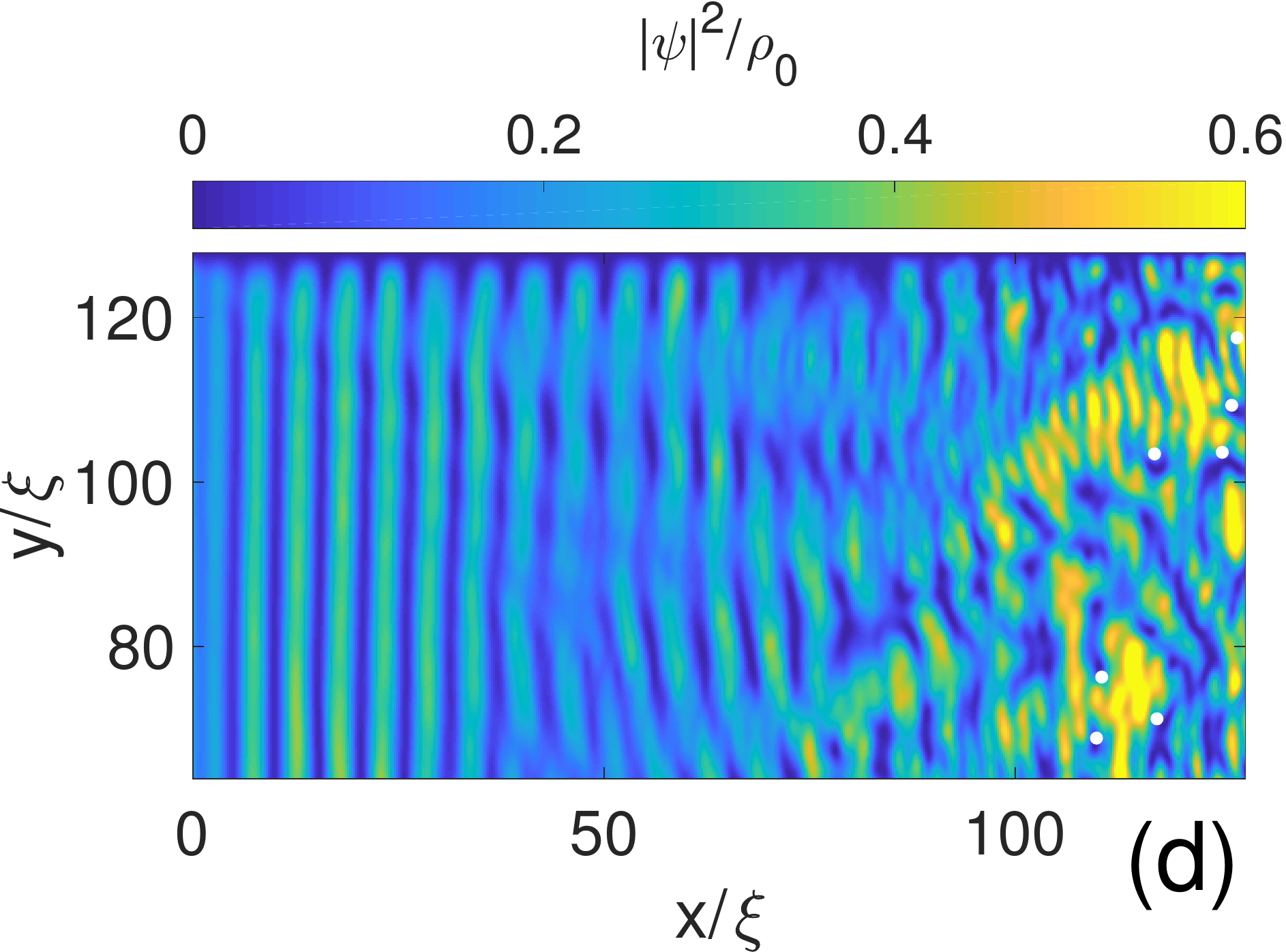}\\
\includegraphics[width=0.24\linewidth]{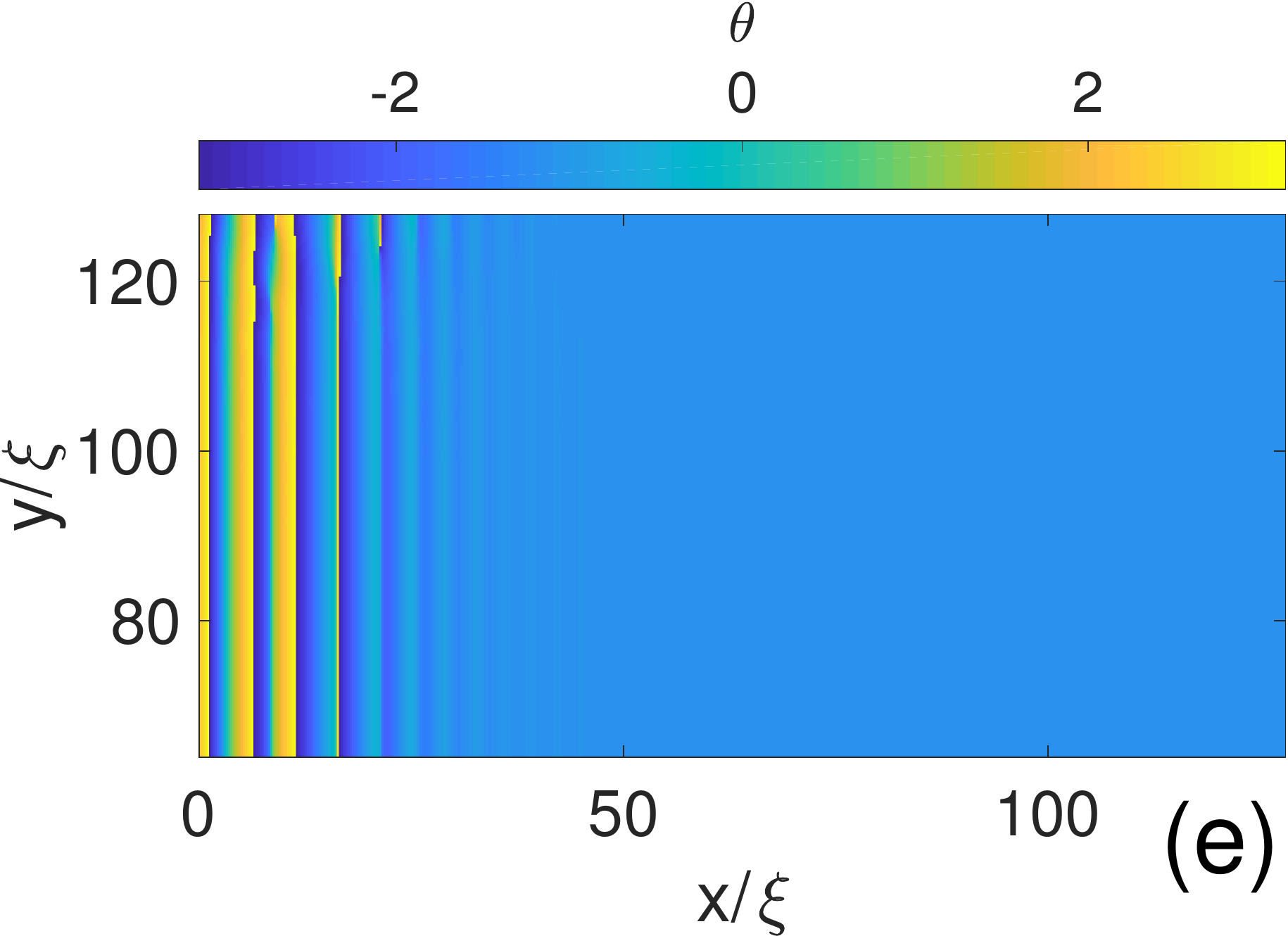}
\includegraphics[width=0.24\linewidth]{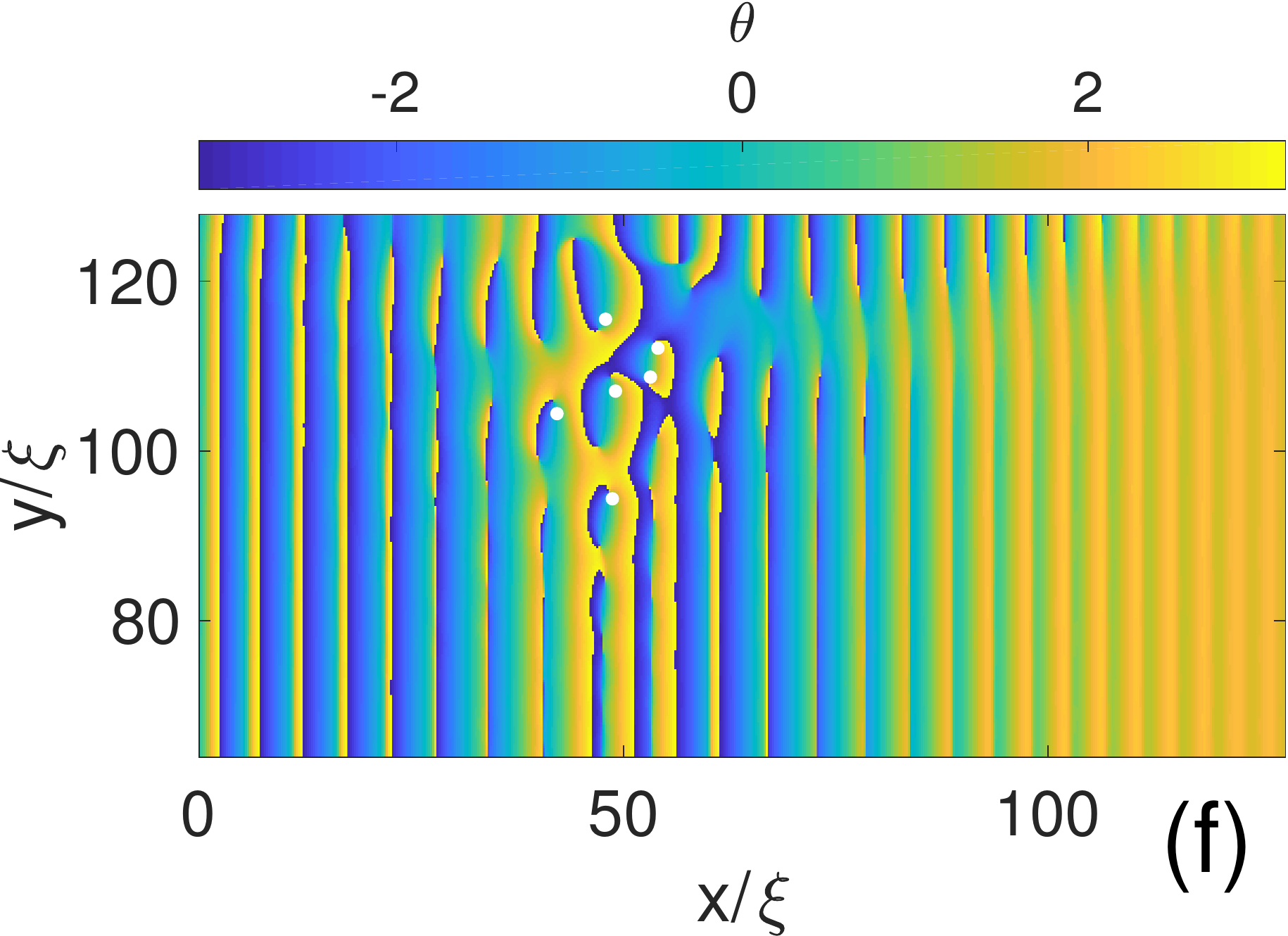}
\includegraphics[width=0.24\linewidth]{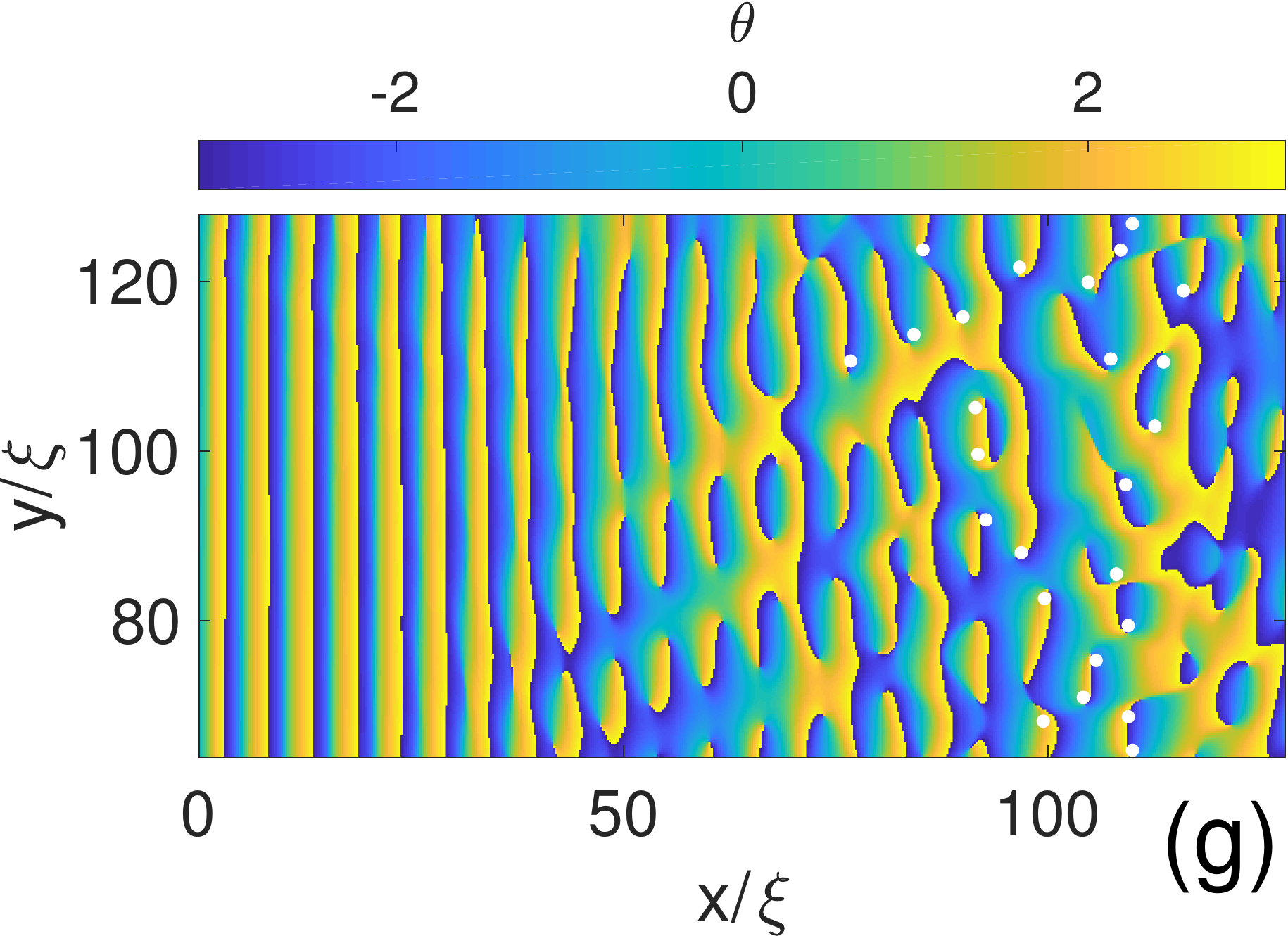}
\includegraphics[width=0.24\linewidth]{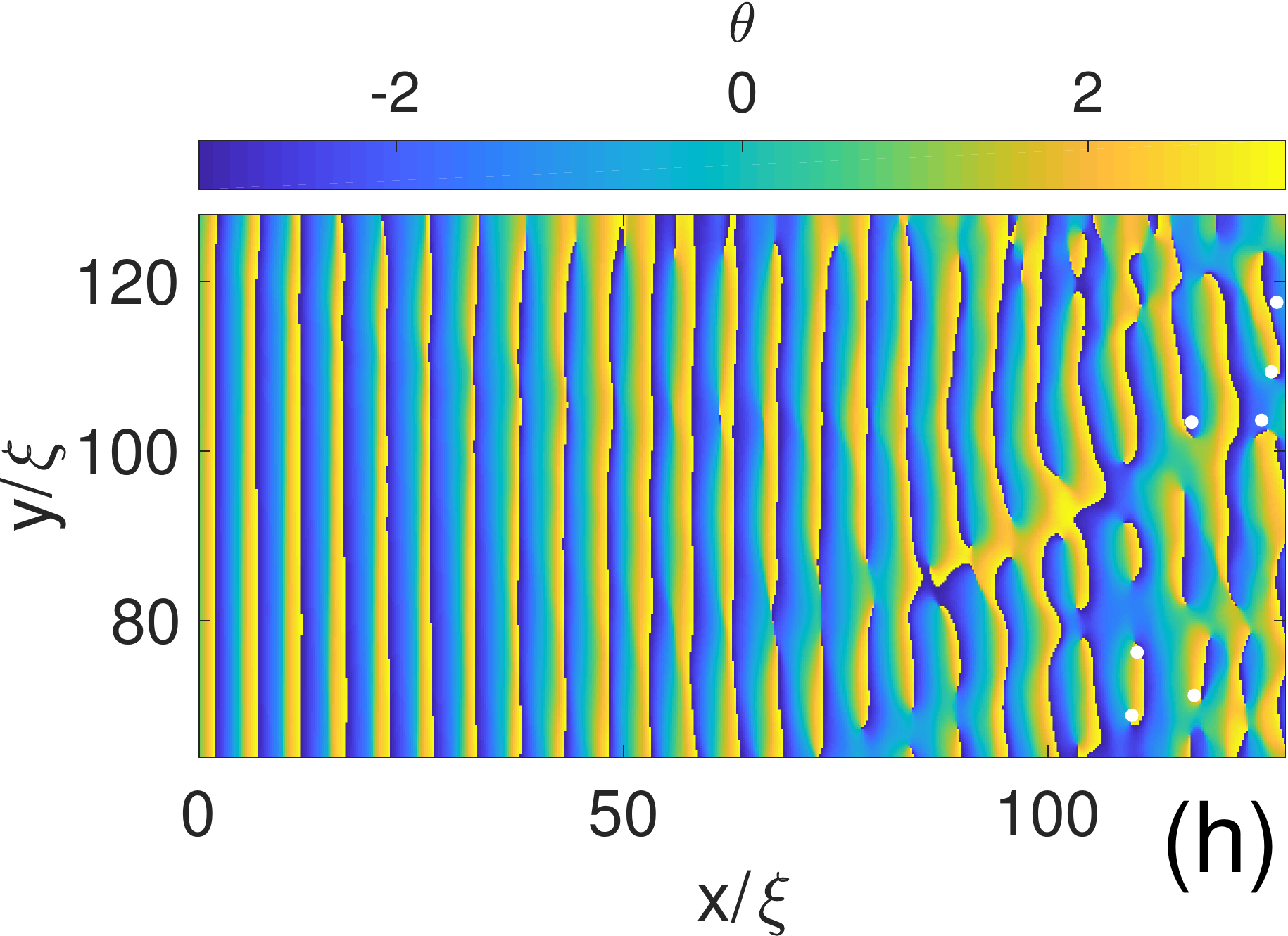}
\caption{{\it (Color online)} {Zoom on density and phase fields in simulation with $V_0=1.5/\mu$, $\sigma=1.2/\xi$ and $Z_0= 0.88$}. Panels {(a-d)}: density field, {(e-h)}: phase field. The zoomed window is the red rectangle in Fig.~\ref{DENSL}(a). Panels (a,e): $t=30\xi/c$,  (b,f): $t=60\xi/c$, (c,g): $t=100\xi/c$, (d,h): $t=150\xi/c$.}\label{SOLS}
 \end{figure}

In order to highlight the dynamics of the snake instability, we present an example of the early stage 2D density and phase fields with potential strength $V_0 = 1.5/\mu$ and initial imbalance $Z_0 = 0.88$. We zoom in on the  subregion of $B_R$ in Fig.~\ref{SOLS} (a), the region depicted by the red rectangle in Fig.~\ref{DENSL} (a). 
The train of quasi 1D solitons within a dispersive shock region is seen in Fig.~\ref{SOLS} {(a)}. Fig.~\ref{SOLS} {(b)} shows the snake instability forming along the solitons.  As they travel, the solitons begin to snake until they break up into a chain of ghost vortices with circulations of alternating signs, which then interact to form vortex turbulence. The ghost vortices can be seen in the corresponding phase plots. For instance, in Fig.~\ref{SOLS} (f) we see discontinuities in the phase where the phase winds from $-\pi$ to $\pi$ around them. The ghost vortices correspond to the dark (blue) circles in Fig.~\ref{SOLS} (b \& c). They are not counted in the number of vortices or marked by the black and white circles corresponding to the well-formed hydrodynamic vortices.

\begin{figure}[ht]
\centering
\includegraphics[width=0.35\linewidth]{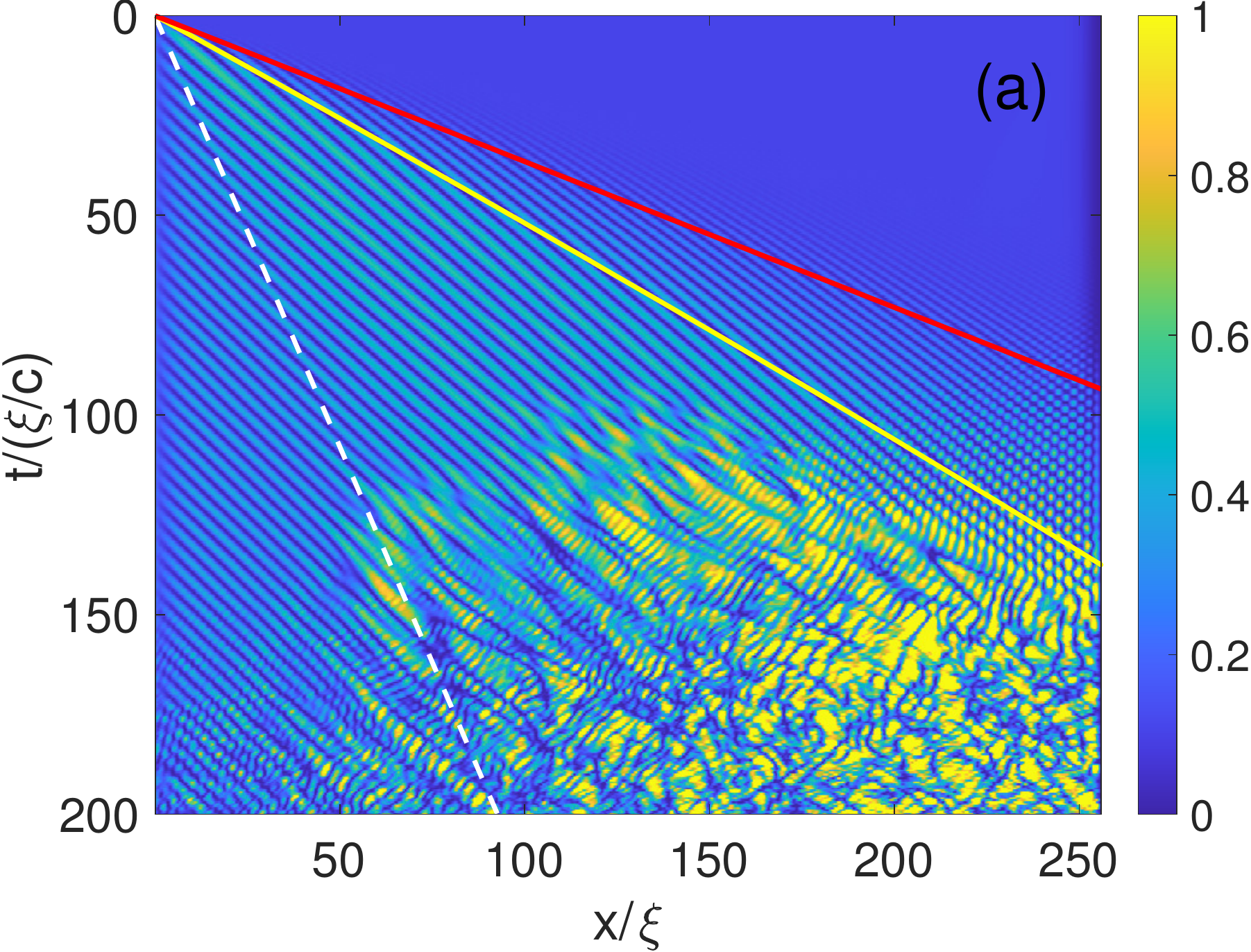}
\includegraphics[width=0.35\linewidth]{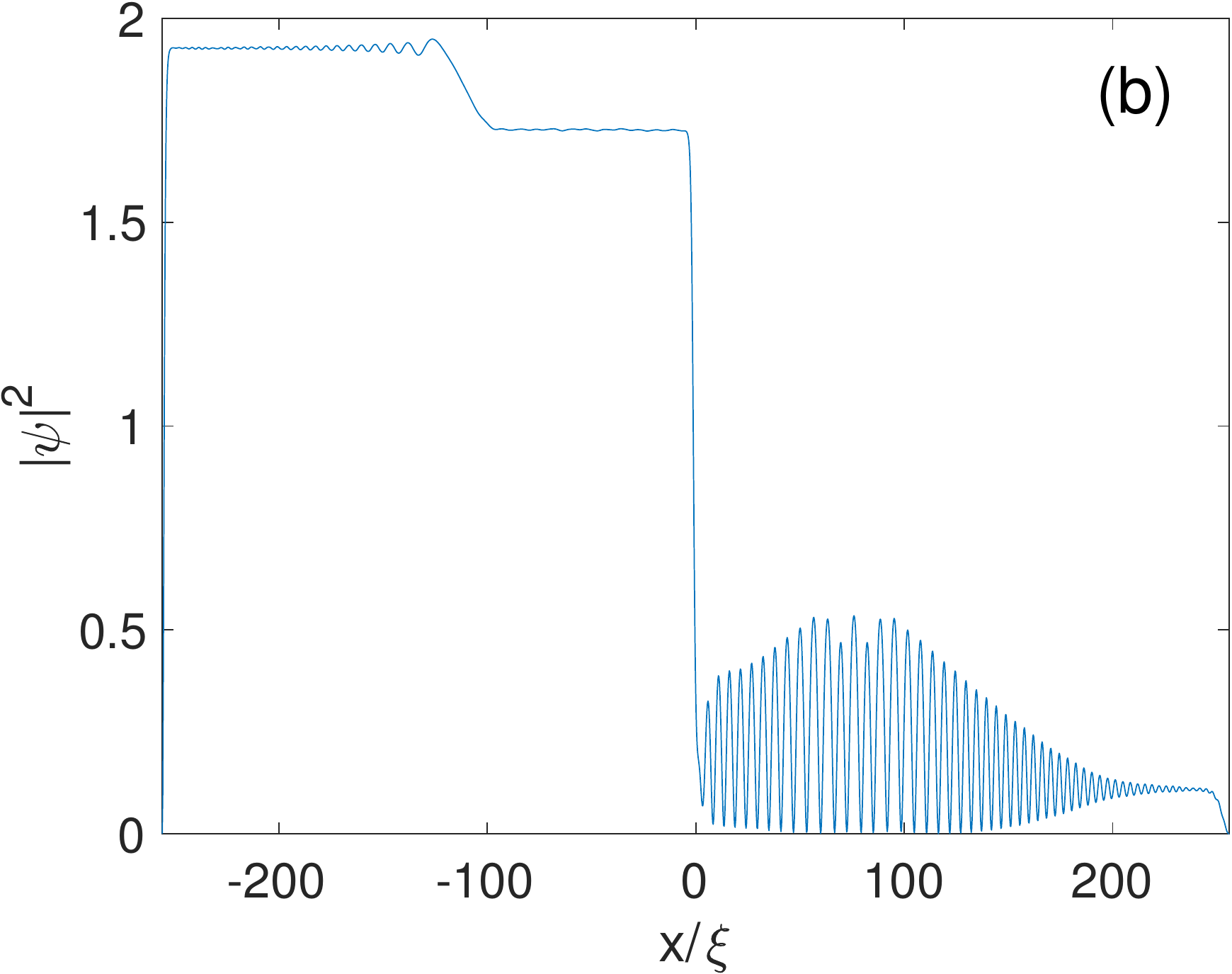}
\caption{{\it (Color online)} Panel (a): slice along positive $x$-axis evolving in time with $V_0=1.5/\mu$, $\sigma=1.2/\xi$ and $Z_0= 0.88$. Overlaid line indication average speed of sound in the left box at $t=0$ travelling away from the origin (yellow), the speed of the front as calculated in \cite{El:1995aa} (red) and the speed of sound in the right box at $t=0$ (dashed white). Panel (b): slice for entire $x$-axis at the time $t = 8\xi/c$.}\label{Slice}
 \end{figure}

The speed of the oscillatory front (of the train of solitons) of the dissipative shock wave is calculated in \cite{El:1995aa} for the 1D case. In Fig.~\ref{Slice}(a) we show that our 2D simulations have a good agreement with the calculation made in \cite{El:1995aa}. We see that the shock moves faster than the local speed of sound of even the high density region in $B_L$. The large solitons move at a speed which is faster than the initial speed of sound in $B_R$, and slower than the initial speed of sound in $B_L$. It is also possible to see the snake instability and subsequent chaos after times of around $t=100\xi/c$. In Fig.~\ref{Slice} (b) we see that there is also a rarefaction pulse moving in the negative x-direction, also in agreement with \cite{El:1995aa}.

In the case of many atoms, where the Laplacian term in the Gross-Pitaevskii equation can be neglected, we can analytically find the stationary profile for the wave function. Such a profile is known as the Thomas-Fermi (TF) profile. We choose to take a slice along $y=0$ as there is no $y$-dependence of the potentials.
The fluid having enough energy to flow over the barrier corresponds to a TF profile such that $|\psi(x)_{TF}|^2>0$ for all $x$. In this case, the dynamics are different to that of the classical Josephson junction as the flow is not only a consequence of tunnelling but also due to the fluid which has enough energy to pass the barrier. The stationary TF profile is given by: 

\[ \begin{cases} 
      \psi_{TF}  = \sqrt{\rho_0-V_0e^{\frac{x^2}{\sigma^2}} - V_d \tanh(x)} \quad \text{if} & \rho_0-V_0e^{\frac{x^2}{\sigma^2}} - V_d \tanh(x) >0, \\
   |\psi_{TF}|^2  =  0 & \text{otherwise.}
   \end{cases}
\]
Thus the barrier TF width, $W_{TF}$, is given by:

\begin{align}
W_{TF}=\sigma\sqrt{\ln\left(\frac{V_0}{\rho_0+V_d}\right)}.
\end{align}
From this we can see that flow has no tunnelling when $V_0<\rho_0+V_d$. 
Once the vortices are introduced, confining them to the right box only will increase interactions due to the smaller inter-vortex distance. Vortices tend to spread out by forming dipoles which move away from the vortex bulk at a nearly constant speed. These vortices can penetrate the barrier in certain cases, namely when $V_0$ is small and/or the vortex dipole is fast. The barrier width $W_{TF}$ will also affect the ability of vortices to penetrate the barrier. As the vortex-barrier interaction is non-linear, it is not clear how the transmission of vortex dipoles is affected by certain parameters, such as $V_0$, $\sigma$ and the size of the vortex dipole. To shed light on this interaction we will now present a study to classify and quantify different outcomes of the dipole-barrier interactions.

\subsection{Vortex dipole scattering off the barrier}\label{VDSWB}
The interaction between quantised vortices and the JJ barrier plays a vital role in the dynamics discussed in section \ref{CVd} and is an interesting problem in itself. The barrier can trap vortices as well as assist in their annihilation. 
In certain regimes of 2D vortex turbulence, vortices tend to couple into vortex dipoles \cite{Nazarenko_2011}. This process is the result of random vortex motion; it includes inter-vortex collisions that can re-couple or scatter vortex pairs. As a result of this motion some vortex pairs will move away from the turbulent bulk. As we have boundaries, the vortices will be incident either on the outer boundaries or the barrier. If the dipole is incident on the outer boundary, it will get split into the two vortices  moving along the boundary in opposite directions. If the dipole is incident on the barrier, it can be transmitted, annihilated or trapped, depending on the barrier height and the dipole size.

\begin{figure}[ht!]
\centering
\includegraphics[width=0.23\linewidth]{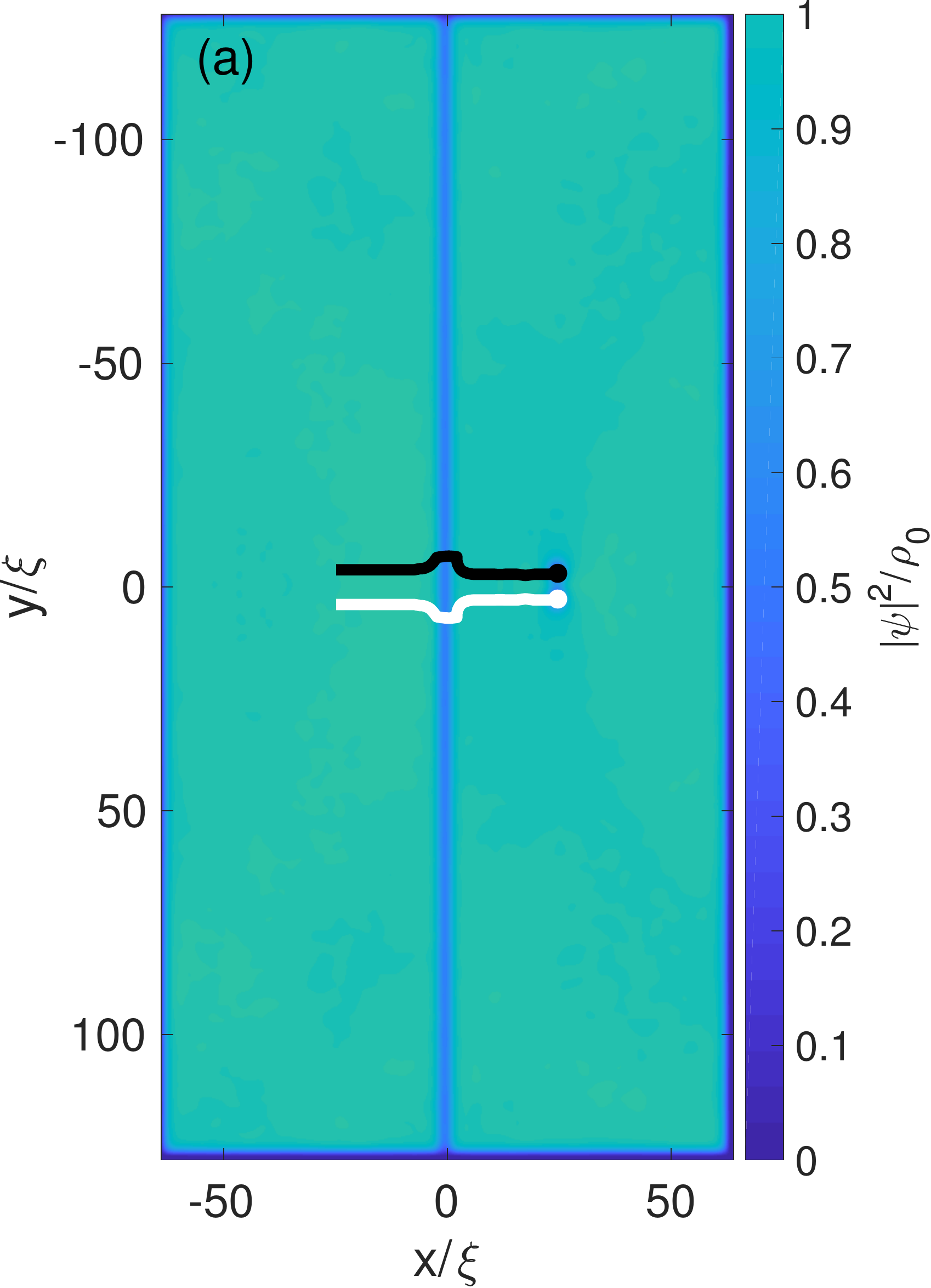}
\includegraphics[width=0.23\linewidth]{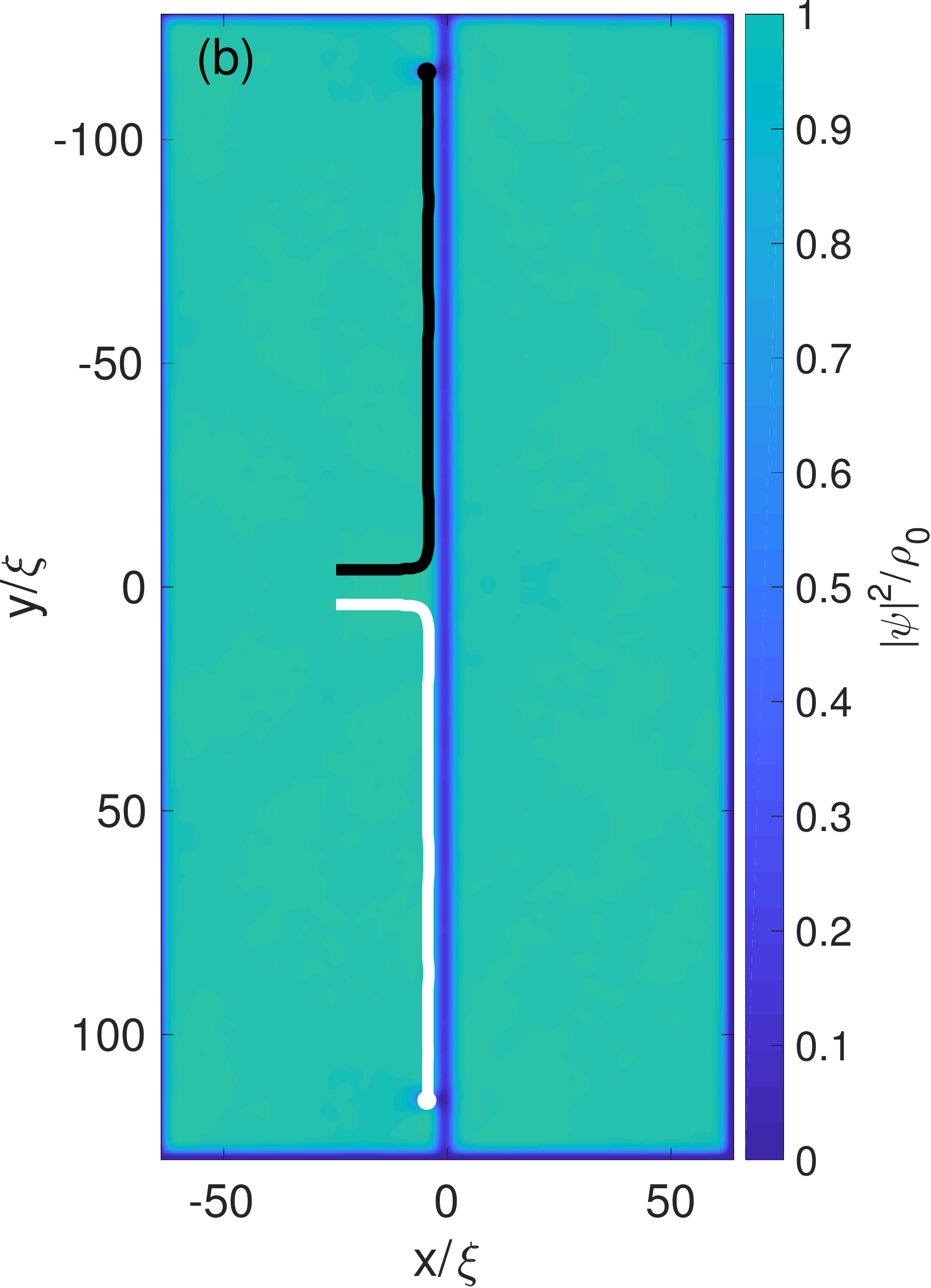}
\includegraphics[width=0.23\linewidth]{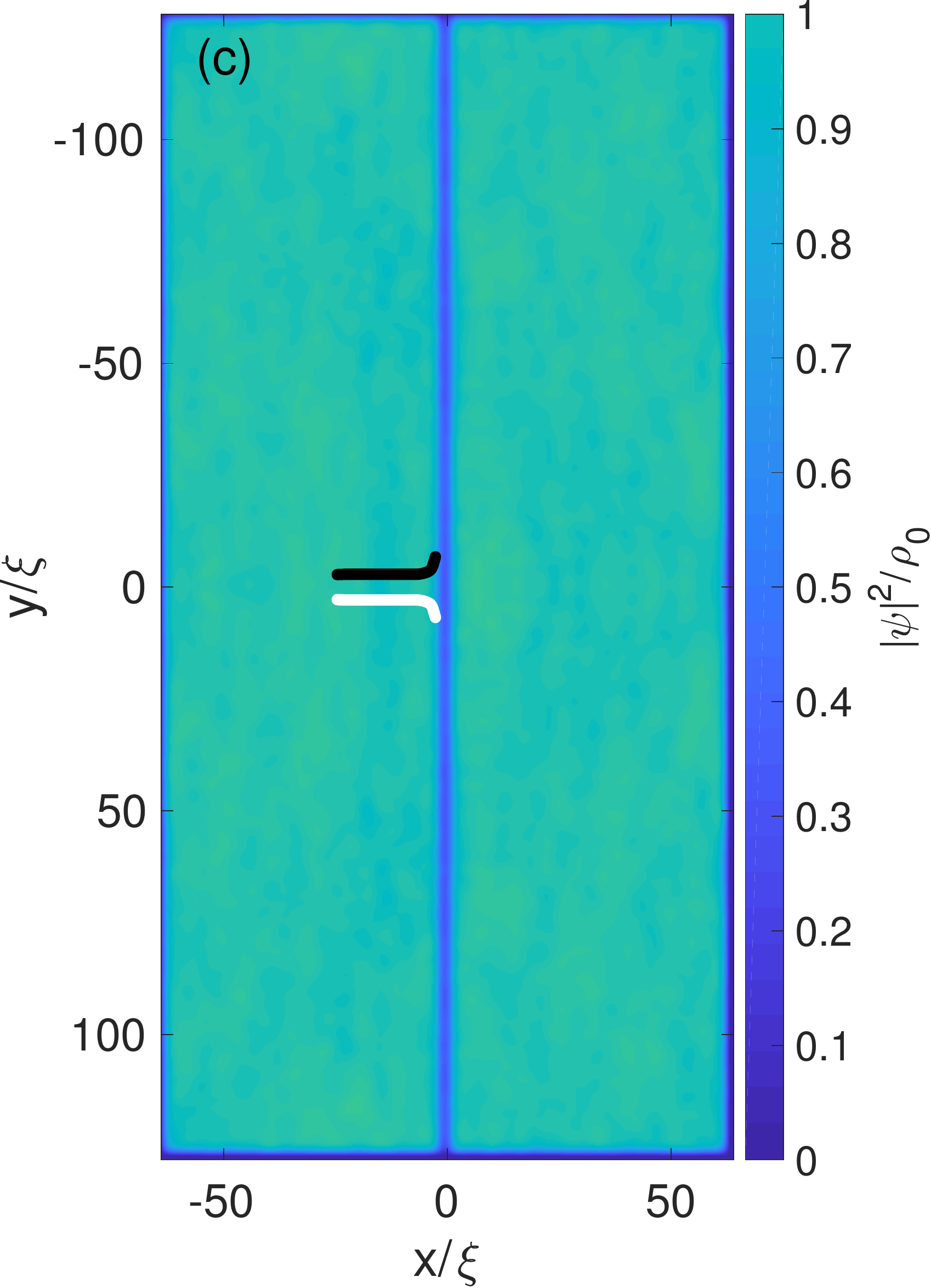}
\includegraphics[width=0.23\linewidth]{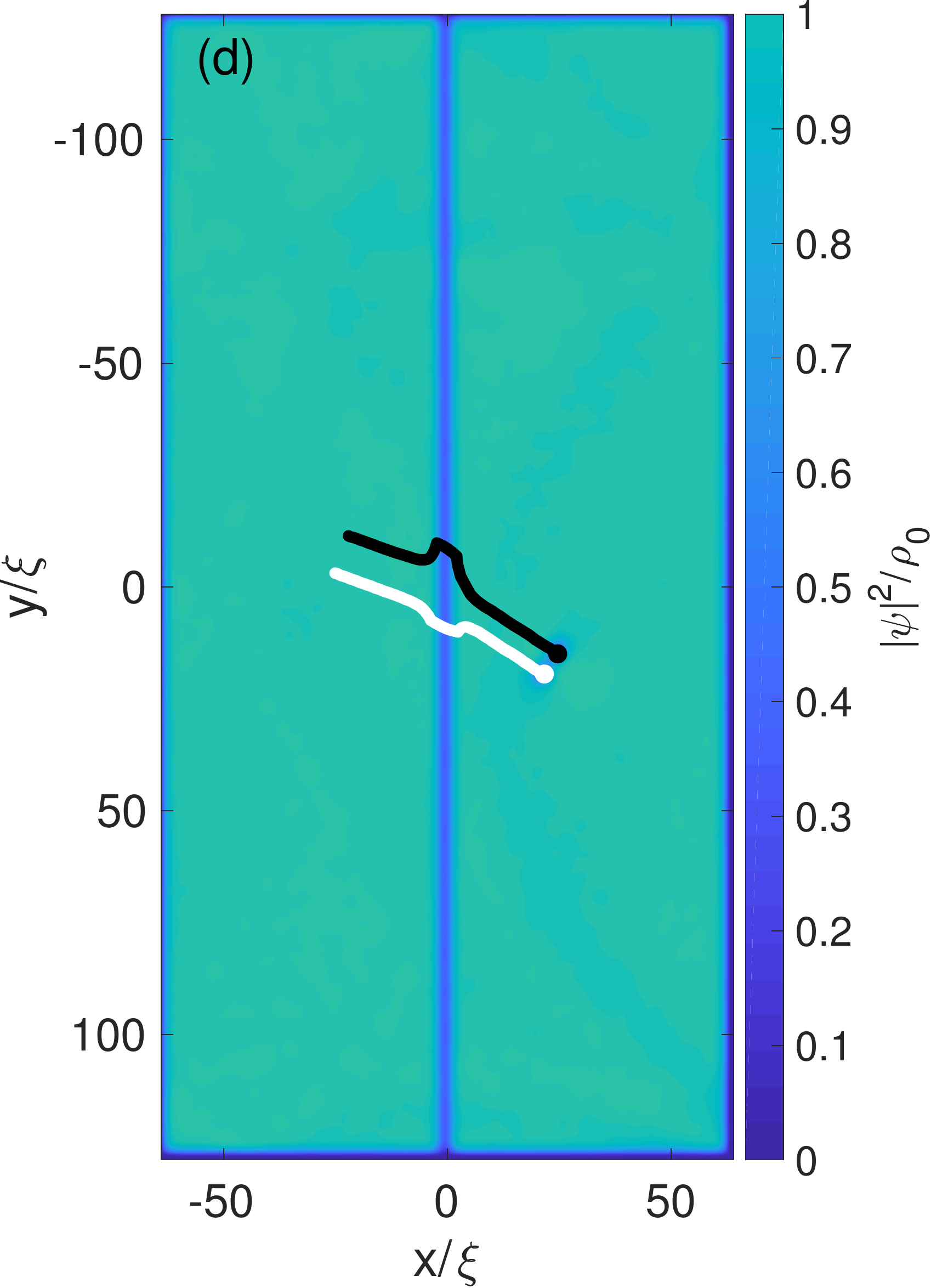}
 \caption{{\it (Color online)} Examples of vortex-barrier interactions. Panel (a): Vortices crossing the barrier at $V_0 = 0.8/\mu$ and $d_0 = 8/\xi$. Panel (b): Vortices trapped by the barrier at $V_0 = 1.9/\mu$ and $d_0 = 8/\xi$. Panel (c): Vortices annihalting $V_0 = 1.2/\mu $ $d_0 = 6/\xi $; (d): Vortices passing with an initial angle of incidence of $0.4 \pi$, $V_0 = 1.15/\mu $ and $d_0 = 10/\xi $. The white line corresponds to the vortex trajectory and the black line corresponds to the antivortex. }\label{DNO}
\end{figure}

In this subsection we present a study of a vortex dipole interacting with a JJ barrier where no initial density imbalance between the left and right boxes is present, that is, $Z_0=0$ and $V_d=0$. 
Initially, we position a vortex dipole centred at $(-25\xi,0)$, and define $\theta$ as the angle between the $ x $-axis and the direction of propagation of the vortex dipole. Clearly by symmetry, we expect the dynamics to be mirror-symmetric with respect to the $x$-axis i.e. with respect to the change $\theta \to -\theta$. 
The vortices are initially separated by a distance of $d_0$, and the vortex with positive circulation is in the upper-half plane so that the dipole moves towards the positive $x$-direction. Examples of the resulting vortex trajectories are shown in the different panels of Fig.~\ref{DNO}. In Fig.~\ref{DNO} we see that in all examples the vortices lose energy to sound during the interaction with the barrier. We quantify the amount of sound emitted by calculating the change in the incompressible energy $$\Delta E^{i}(t)=E^{i}(0)-E^{i}(t) = \int_{B_L+B_R} \left(\varepsilon^{i}_{kin}(x,y,0) -\varepsilon^{i}_{kin}(x,y,t) \right)dxdy.$$
We measure this change in energy at a final time $t_f$ after the interaction has happened. As a criteria for determining $t_f$ we choose one of the following three criteria: (i) the vortex dipole passed the line $x=25\xi$, corresponding to Fig.~\ref{DNO} (a,d); (ii) either vortex becomes within $5\xi$ of the system boundary, corresponding to Fig.~\ref{DNO} (b); (iii) if case (i) and (ii) are not fulfilled we allow a maximum time of $ t_f =750 \xi/c$. We assume that the vortices have annihilated if they do not fulfil case (i) or (ii) after such a long time, so case (iii) corresponds to vortex annihilations; an example is shown in Fig.~\ref{DNO} (c).

Fig.~\ref{DNO} shows four different examples of the dipole-barrier scattering for different values of the scattering parameters $ V_0 $, $ d_0 $ and $ \theta $.
The images show the superfluid density plots after the scattering with the dipole with the trajectories overlaid. For the dipole-barrier scattering experiments presented in this subsection we have chosen to shorten the $ L_x $ side and increase the $ L_y $ side of the JJ system, compared with the one presented for instance in Fig. \ref{DENSL}, in order to give the vortices more space to interact with the barrier. Fig.~\ref{DNO} (a) shows the path two vortices take when passing a barrier. Notice that the vortices after the interaction are closer together. As the vortices' motion was perpendicular to the barrier there was no deflection. Fig.~\ref{DNO} (b) shows the case of a dipole not being able to pass the barrier, this corresponds to case (ii). The vortices separate from one another and move along the barrier with the vortices effectively see images of themselves in the barrier. This motion is similar to that when a vortex pair is incident on an external boundary, where the boundary conditions are similar to that of an infinite barrier. Fig.~\ref{DNO} (c) shows an annihilation, this corresponds to case (iii). Fig.~\ref{DNO} (d) shows the vortices being deflected during the interaction with the barrier. In this example the vortices were not moving perpendicular to the barrier but had an incidence angle $\theta=0.4\pi$. 

\begin{figure}[ht]
\centering
\includegraphics[width=0.45\linewidth]{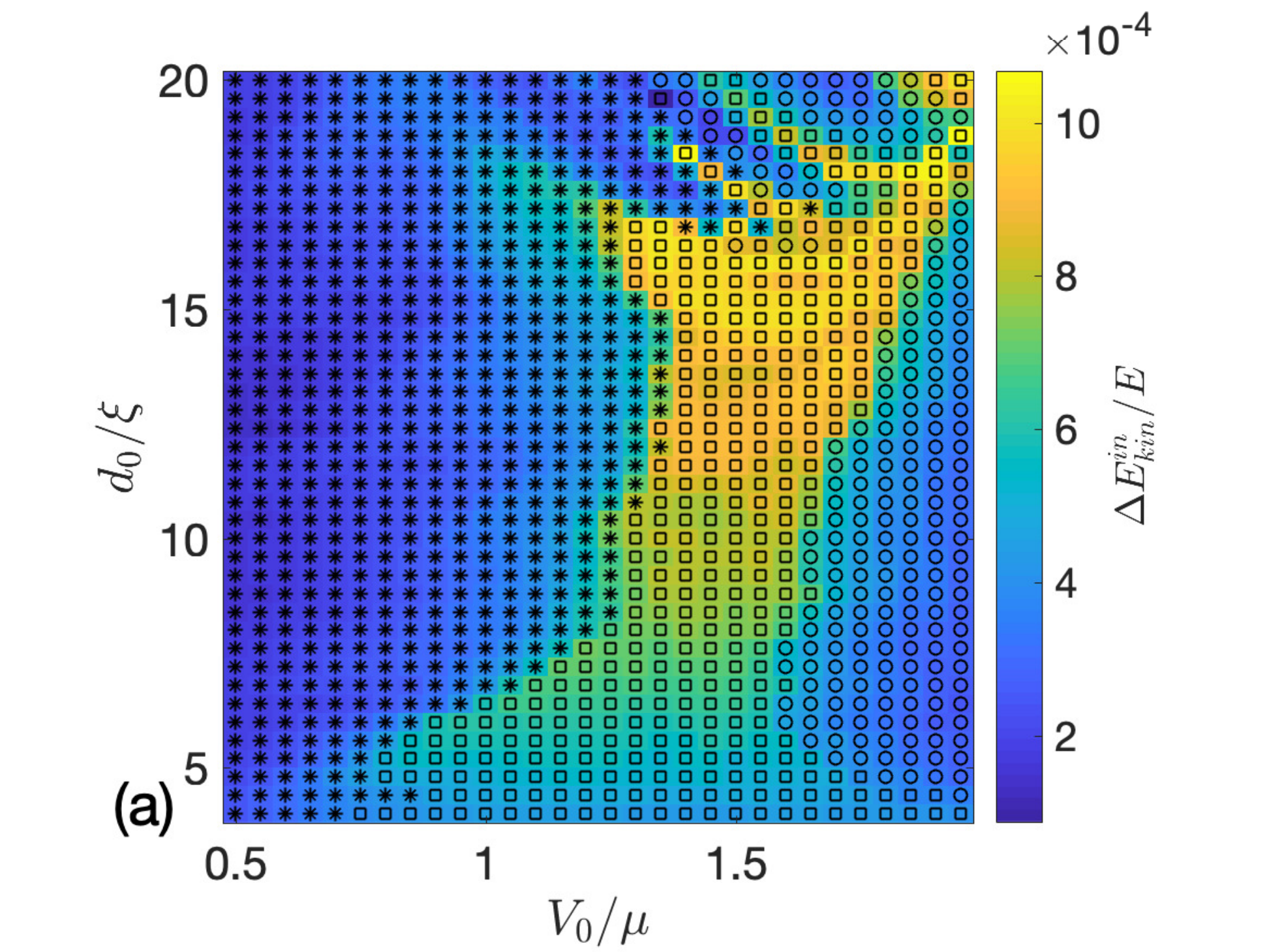}
\includegraphics[width=0.45\linewidth]{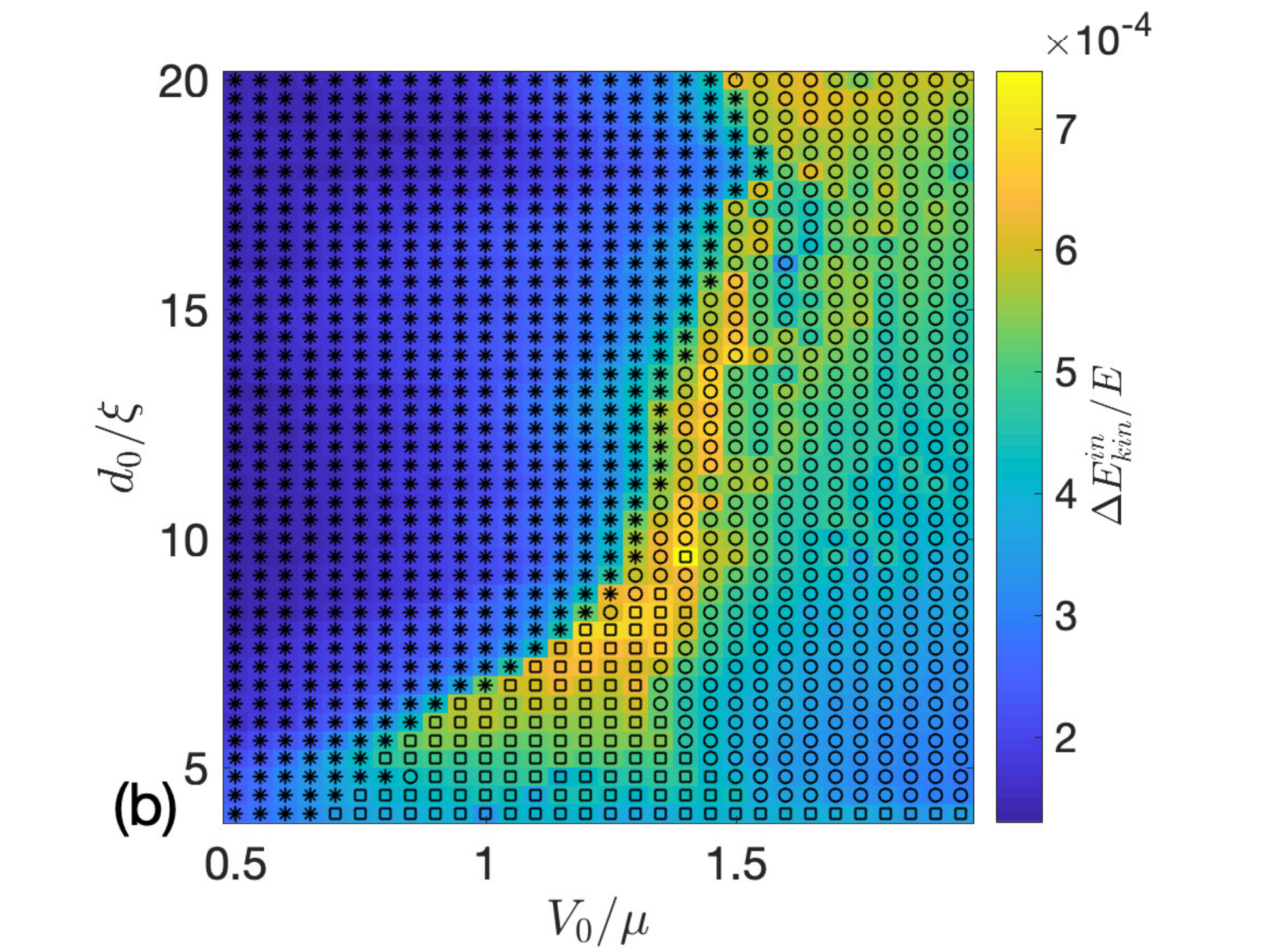}
 \caption{{\it (Color online)} Panel (a): parameter sweep of initial dipole separation and barrier height for angle of incidence $\theta=0$. Scaled colour indicates the approximated compressible energy  at $t=t_f$. Overlaid points show if the dipole passed (asterisks), annihilated (squares) or interacted with boundary (circles). Barrier width $\sigma=1.2$. Panel (b): similar parameter sweep for $\theta = 0.4\pi$. }\label{DVV}
\end{figure}

We perform two sets of simulations for two different incidence angles $\theta = 0, 0.4\pi$, sweeping over different vortex separations $d_0$, and vortex barrier heights $V_0$. We observe and classify for each set of parameters what kind of dipole barrier interaction takes place, (i), (ii) or (iii), and also measure the sound released in the interaction. The results are presented in Fig.~\ref{DVV}. Firstly we calculate whether or not a vortex dipole of separation $d_0$ will pass the JJ barrier of given strength $V_0$; the cases when the dipoles pass are marked with asterisks. 

If the dipole cannot pass, this can be due to two reasons: the dipole is annihilated by the barrier producing sound (case (iii) marked by squares) or the interaction of the dipole and the barrier is not over when the boundary effects become relevant (case (ii) marked by circles). In the latter case, it is not possible to say whether the dipole would have annihilated in an infinite domain or not, hence it is a consequence of the finiteness of the system. We see that when annihilations happen, there is more sound energy  (measured by the change in the incompressible energy) released from dipoles with larger separations. Fig.~\ref{DVV} shows that there are clear connected regions in which each of the possible cases happens.

As we mentioned in the introduction our aim is to explore the best parameters for vortex turbulence. Annihilations are key events which determine the vortex decay rate and will be further discussed in section \ref{CVd}. We see that the barrier causes annihilations. In Fig.~\ref{DVV} (a) we see that the annihilations are numerous and they also correspond to a high emission of sound. As well as removing vortices from the system, the extra sound is known to increase the vortex decay rate \cite{Nazarenko:2007aa}. On the other hand, the highly energetic sound produced can penetrate the barrier and be spread over the adjacent box which is void of vortices. As a result, the sound is distributed over twice the area (both the wells). Thus, the second well acts as a sound absorber i.e. as an effective heat sink. By introducing a finite angle of incidence $\theta$, the amount of vortex annihilations is greatly reduced. At the same time vortex splitting becomes much more frequent. However, the amount of vortices that pass the barrier does not change much. We see that small dipoles ($d_0<7$) are annihilated by the barrier in the region of the barrier strength $V_0 \sim 1$. For the region $V_0= 1.3-1.6 $ with large dipoles ($d_0>7$) we see that the dipoles separate. Compared to the splitting region in Fig.~\ref{DVV}(a) in this case they emit much more sound.


The vortex-barrier interaction discussed here is more simple than the interaction when the densities are unequal, the background condensate is saturated with sound, and there are more than two vortices involved in the interaction. The results are informative as a rough measure of the vortices ability to cross the barrier. Similar studies of dipoles incident on a sharp density gradient have been under taken in \cite{Cawte:2019aa}, where the authors discover a dependence of the scattering angle similar to a Snell's law.

\subsection{Vortex turbulence}
\subsubsection{Optimal parameters}
We now discuss the optimal choice of parameters to produce vortex turbulence. Our aim is not only to produce the highest number of vortices, but to also minimise the secondary by-products of the method, namely, sound and large density waves.
We introduce the mean number of vortices over time

\begin{align}
\bar{N}_v = \frac{1}{T}\int^{T}_0 N_v(t) dt,
\end{align}
and for the left and right box, $\bar{N}_{V_L}$ and $\bar{N}_{V_R}$ respectively. We use the mean (opposed to the maximum) as a measure as it also takes into account the sustainability of the vortex turbulence. We not only want to create many vortices, we also want them to persist for as long as possible. Another measure we use to classify the quality of the turbulence is the amount of interfering compressible waves, these account for the large scale density sloshing and the small scale acoustic component. Obviously, it is desirable to minimise such compressible motion.

\begin{figure}[ht]
\centering
\includegraphics[width=0.3\linewidth]{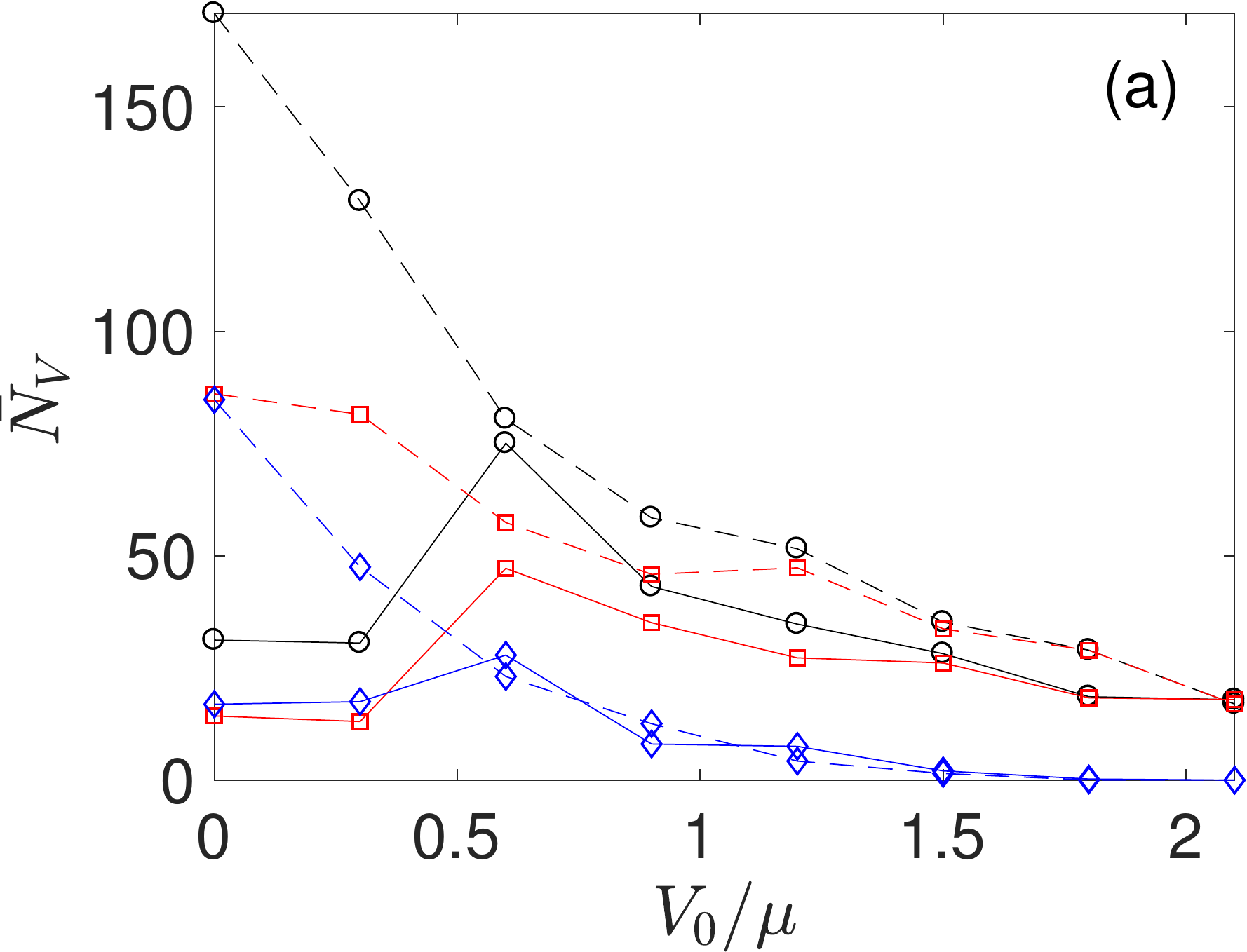}
\includegraphics[width=0.3\linewidth]{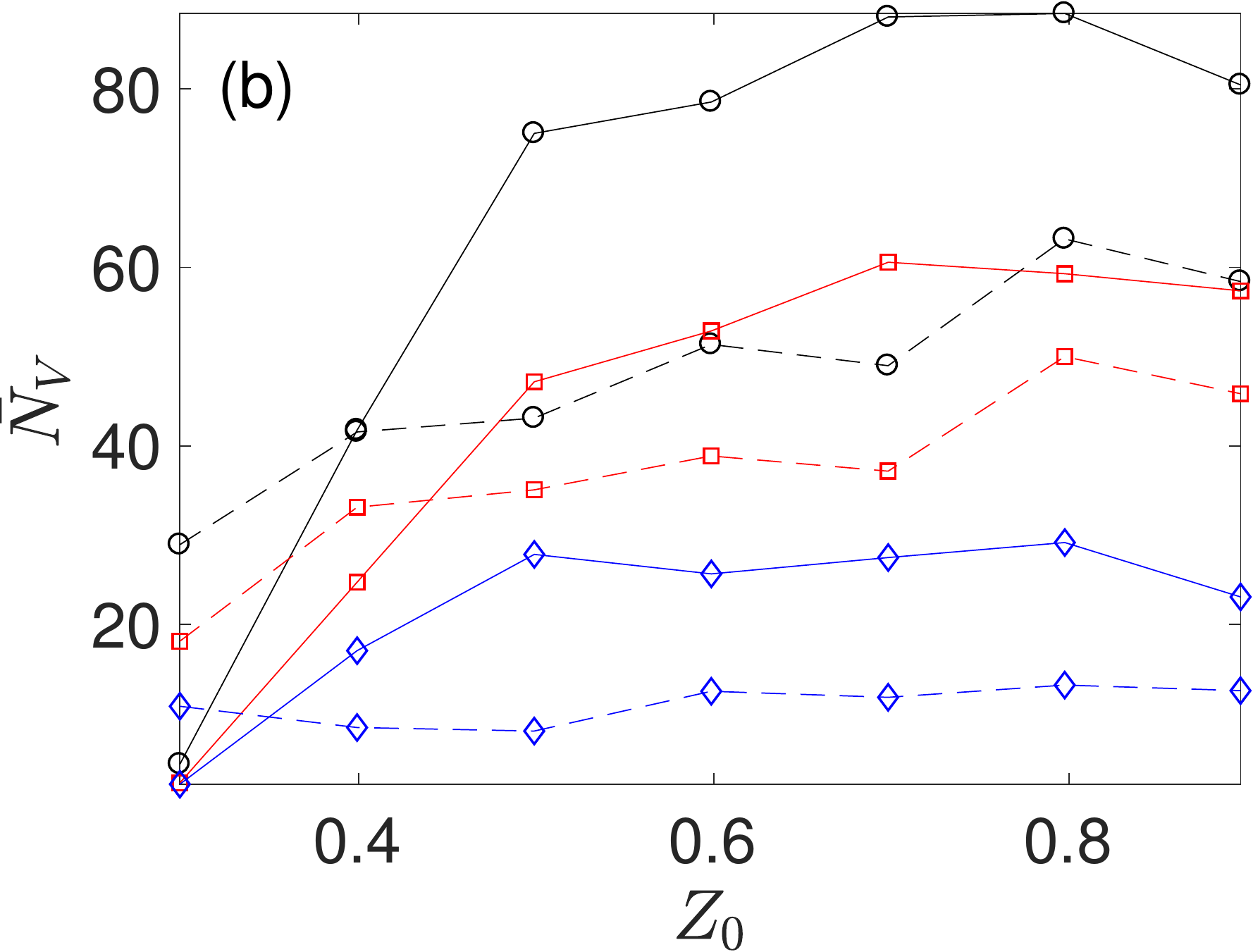}
\includegraphics[width=0.3\linewidth]{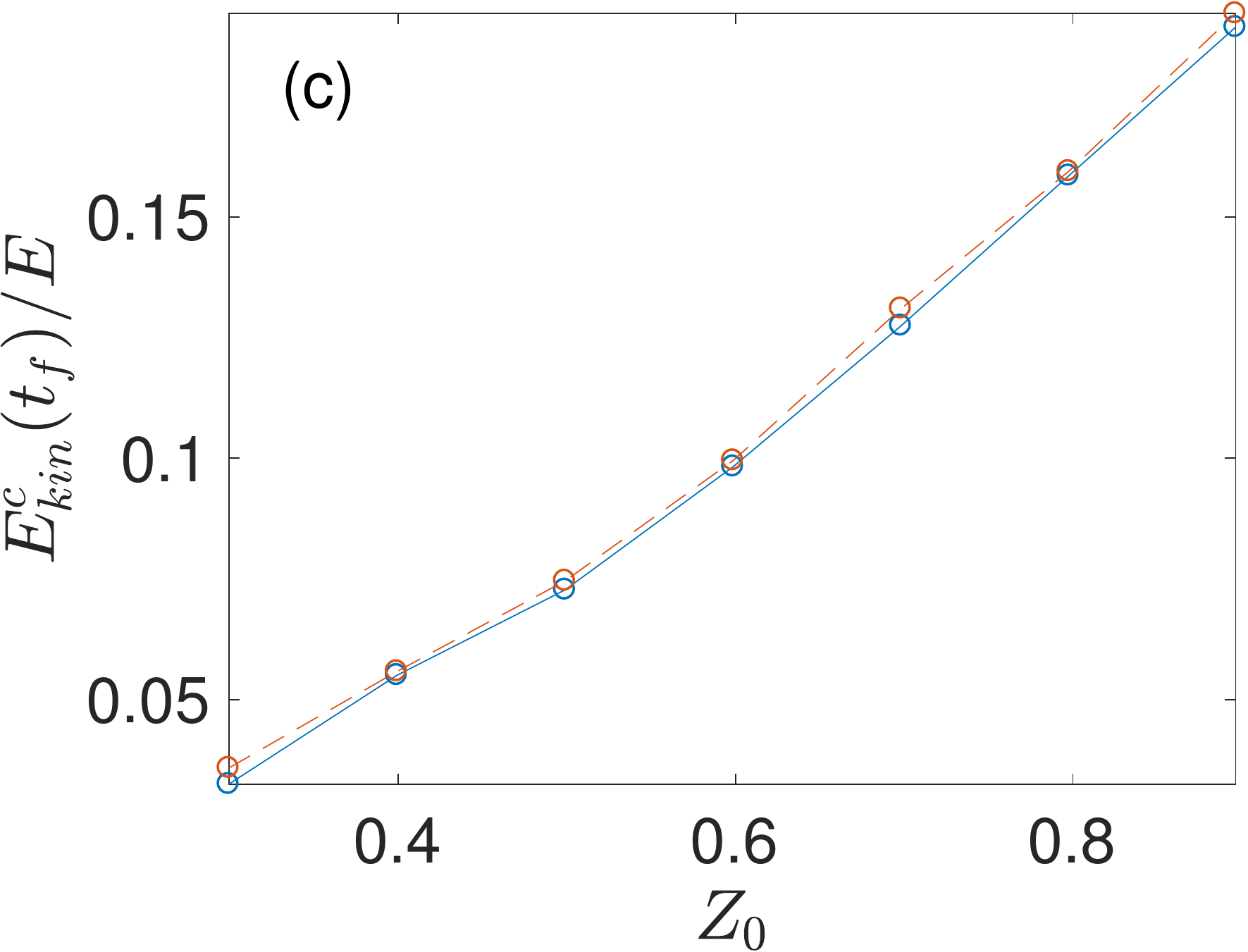}
 \caption{{\it (colour online)} Panel (a): the mean number of vortices against barrier height for two values of initial imbalance $Z_0=0.49$ (full line) and $0.88$ (dashed line). The circle (black), square (red) and diamond (blue) correspond to vortices in $B_L+B_R$, $B_R$ and $B_L$ respectively. Panel (b): how the number of vortices changes for increasing $Z_0$ for two values of barrier height $V_0=0.6/\mu$ (full line) and $0.9/\mu$ (dashed line) for the entire domain $B_L+B_R$ (black circles), the right box only $B_R$ (red squares) and the left box $B_L$ (blue diamonds). Panel (c): the compressible energy as a percentage of the total energy against the initial imbalance for two barrier strengths $V_0 = 0.6/\mu$ (full line) and $0.9/\mu$ (dashed line).}\label{NV}
\end{figure}

The mean number of vortices produced by the proposed method depends both on the barrier strength and the initial imbalance. In Fig.~\ref{NV} we show the effects of varying the two control parameters: the barrier strength $V_0$ and the initial imbalance $Z_0$. In Fig.~\ref{NV} (a) we see that for a large imbalance ($Z_0=0.88$) increasing $V_0$ monotonically reduces the mean number of vortices, and for a large enough $V_0$ no vortices will be produced; this is due to the barrier disrupting the creation of solitons by reducing the rate at which fluid can cross the barrier. On the other hand, for a lower initial imbalance ($Z_0=0.49$), Fig.~\ref{NV} (a) shows that there is a clear maximum for the mean number of vortices in $B_R$ at  around $V_0/\mu=0.6$. This is due to the mechanism discussed in section \ref{INTRO}: a steady influx of density `freezes' the vortices in the right well and the potential reduces oscillations of the density imbalance which would otherwise strongly interact with the vortices. For a high initial imbalance ($Z_0=0.88$), the mean number of vortices is higher without a barrier as shown in Figs.~\ref{NV} (a). However, as we will discuss below, the vortices are accompanied by more small-scale (acoustic) and large scale (sloshing) compressible waves, which is an undesirable effect if we want `clean' vortex turbulence. Even when the initial imbalance is high in Fig.~\ref{NV} (a) and there is no barrier $(V_0=0)$, the vortices are evenly distributed over both boxes. As a consequence the amount of vortices in the right box $B_R$, is not so much larger than the cases with higher values of $V_0$.

When $V_0$ is fixed, by increasing the initial imbalance we reach a plateau in the mean number of vortices in Fig.~\ref{NV} (b). This plateau, along with the increase in compressible energy in Fig.~\ref{NV} (c), indicates that after a certain imbalance the energy is more swiftly converted into compressible sound waves. Using this insight, we propose that it is preferable to choose a lower imbalance such that the vortex dynamics are cleaner, that is, there is less acoustic turbulence and large-scale density sloshing interacting with the vortices.

 A secondary effect also complements the longevity of the vortex turbulence. The highly-energetic small scale sound, produced during the vortex creation and subsequent interactions, can easily pass the barrier. The sound energy from the vortex turbulence of the right well is then distributed over twice the area, reducing the interaction with the vortices and, therefore, slowing down their annihilation rate.

\begin{figure}[ht]
\centering
\includegraphics[width=0.3\linewidth]{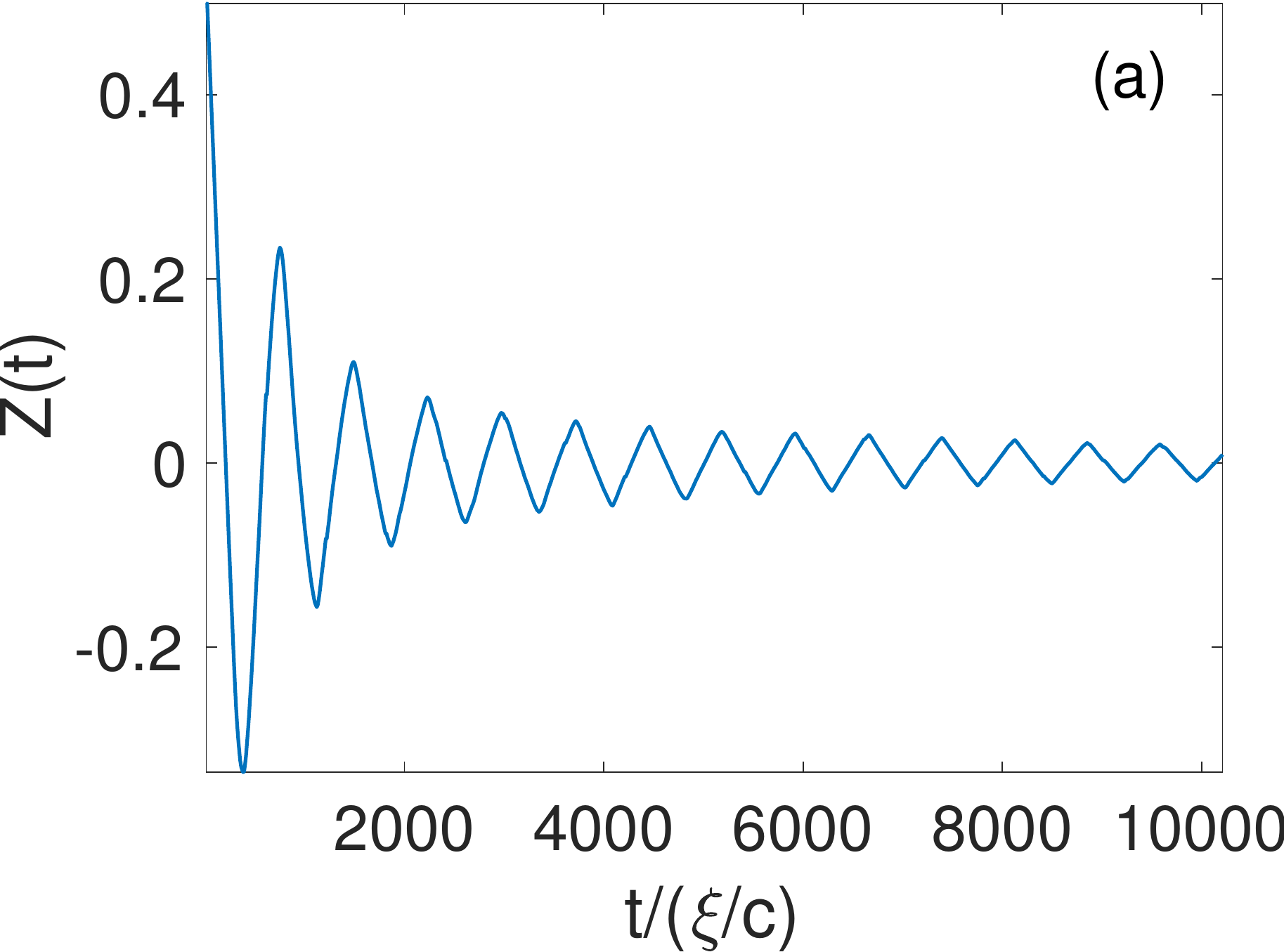}
\includegraphics[width=0.3\linewidth]{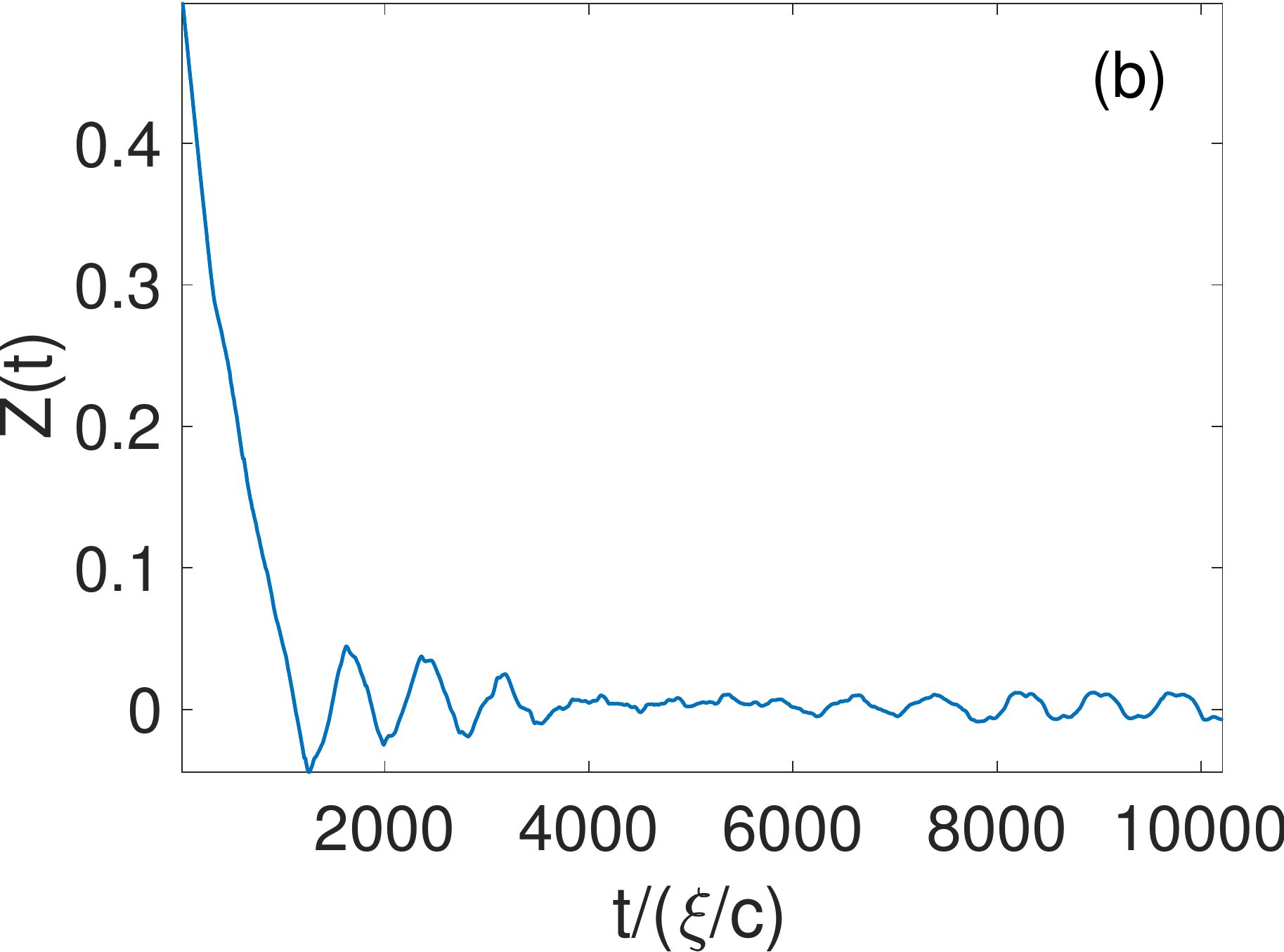}
\includegraphics[width=0.3\linewidth]{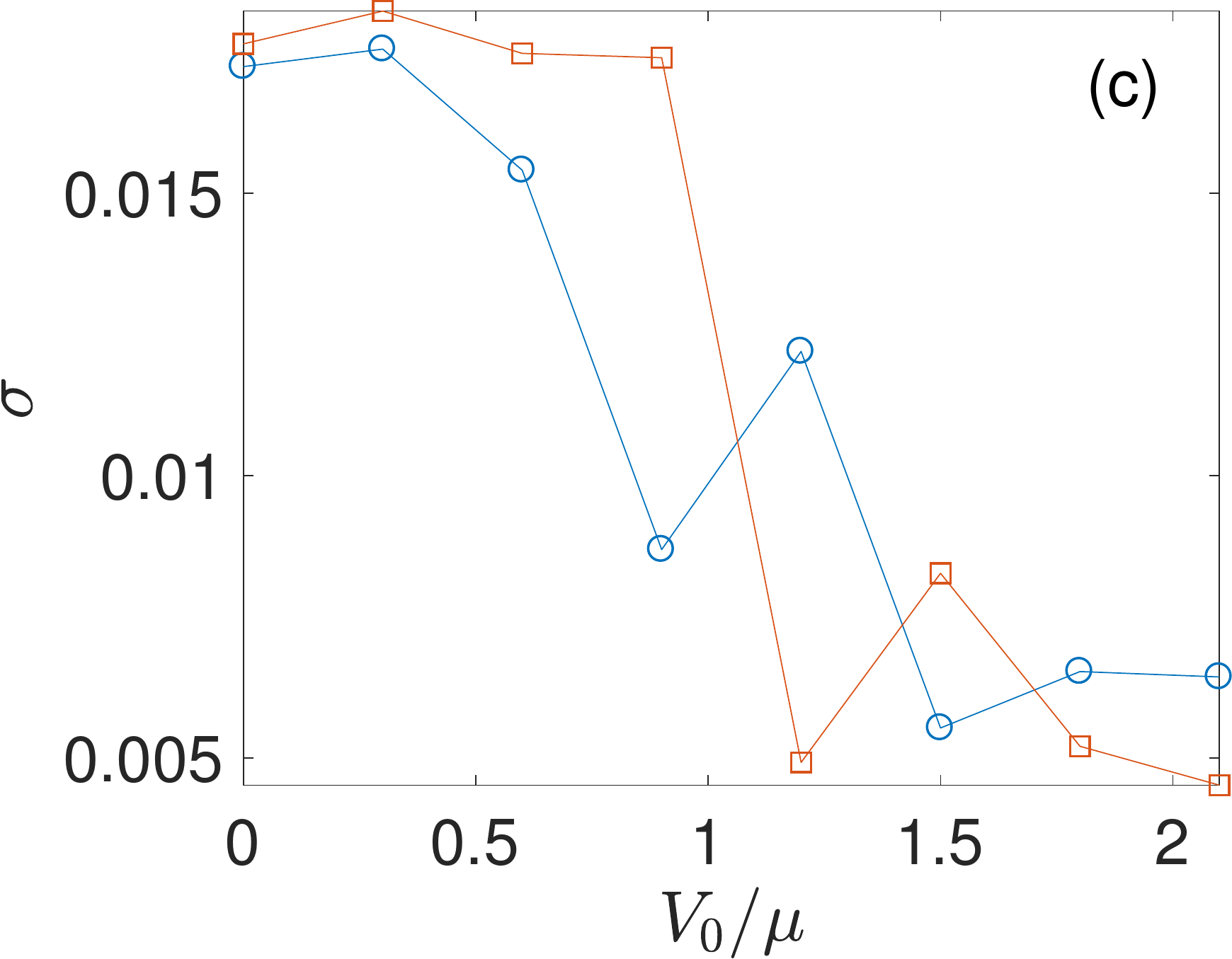}
 \caption{{\it (colour online)} Panels (a): Evolution of relative density $Z(t)$ for parameters $V_0=0.0$, $\sigma=1.2$ and $Z_0 = 0.49$. Panel (b): the same for $V_0=1.5$, $\sigma=1.2$ and $Z_0 = 0.49$. Panel (c): the standard deviation of $Z(t)$  after $t_s=4200 \xi/c$ for a range of different barrier heights. The blue line with circles is for $Z_0=0.49$ and the red with squares $Z_0=0.88$.}\label{ZL}
\end{figure}

In the absence of a barrier (as $V_0 \to 0$), many vortices are produced; however, we identify two critical issues with the turbulence that follows. The first is that the local healing length oscillates (as $Z$ oscillates) for a long time, see Fig.~\ref{ZL} (a). The fluctuation of the local healing length causes vortices to annihilate. Secondly, the large density waves are seen to interact with vortices. This adds additional complexity to the interactions, and it is not clear what the effect it has on the vortex interactions. Introducing a barrier addresses both of these problems as seen in Fig.~\ref{ZL}. When a barrier is present, as in Fig.~\ref{ZL} (b), the oscillations are dampened, with only small oscillations remaining once the boxes have equilibrated. This is due to the amplitude of the large wave simultaneously being reflected and transmitted on each barrier interaction. This multiplies the number of waves and decreases the size of the local wave amplitude. Also, it causes the density to fill the right box more smoothly. The smoother descent is due to barrier reflecting more of the wave when the barrier is stronger. That is, the amplitude of the transmitted wave is reduced on each wave-barrier interaction, therefore, the density flux across the barrier is also reduced; this essentially dampens the overshooting of the oscillations. This can be seen in Fig.~\ref{ZL} (c) where we plot the standard deviation of the oscillations, $\sigma$, against barrier strength: this reduces quickly when the barrier strength increases and a transition seems to occur at a value close to $V_0/\mu=1$. We choose the time period to measure the standard deviation over to be the same for each simulation. The period is chosen to be after all the simulations have reached $Z(t)=0$ for the first time.

\subsubsection{Vortex decay rates}


 Recent discussions \cite{Bagg2018,Cidrim2017,Kwon2014}, indicate that the number of vortices in a homogeneus condensate decay as $t^{-1/3}$, this corresponds to four-vortex interactions. The arguments in \cite{Bagg2018,Cidrim2017,Kwon2014} use a simple logistic equation for the number of vortices $N_v$,
  
  \begin{figure}[ht]
\centering
\includegraphics[width=0.32\linewidth]{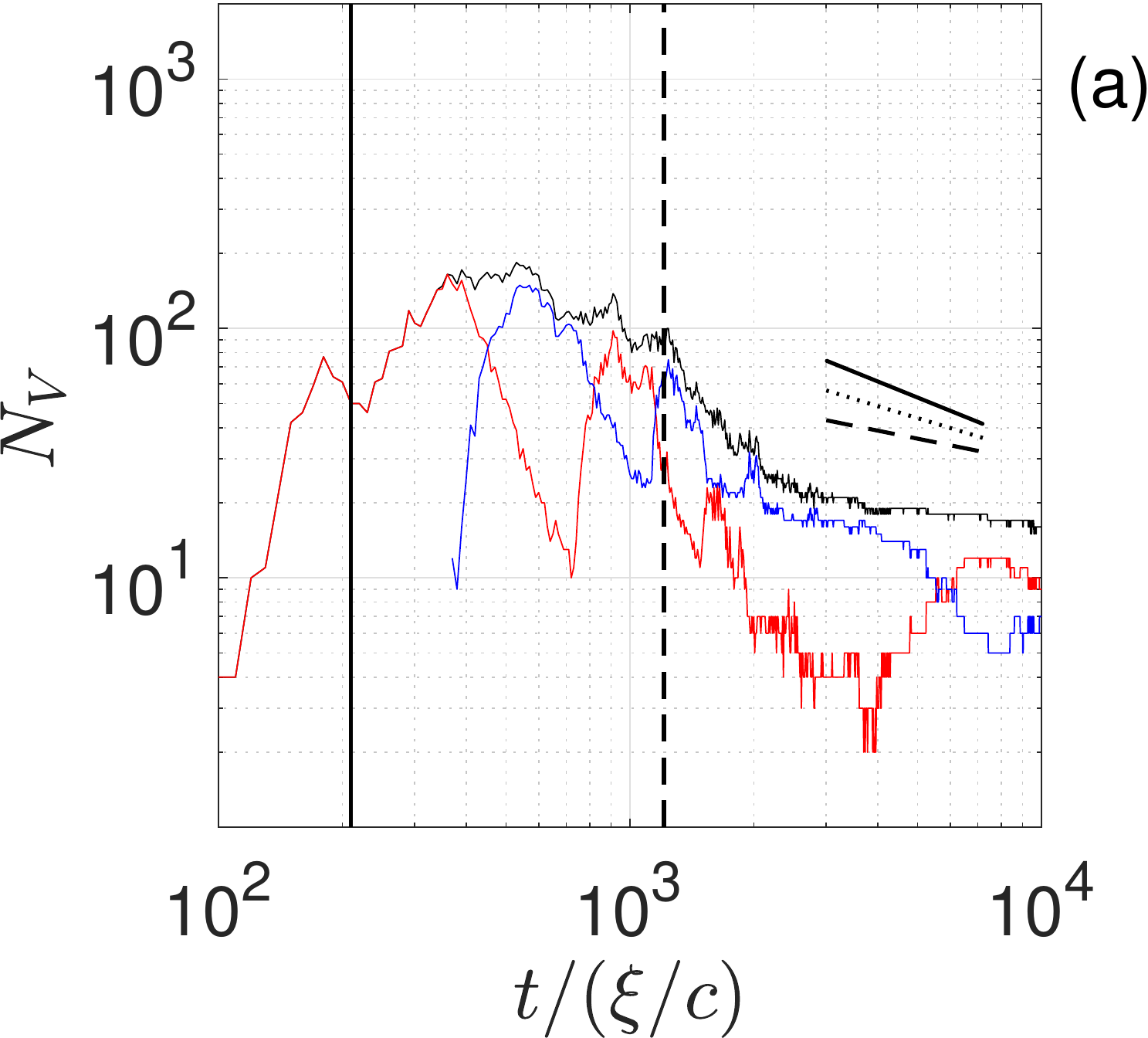}
\includegraphics[width=0.32\linewidth]{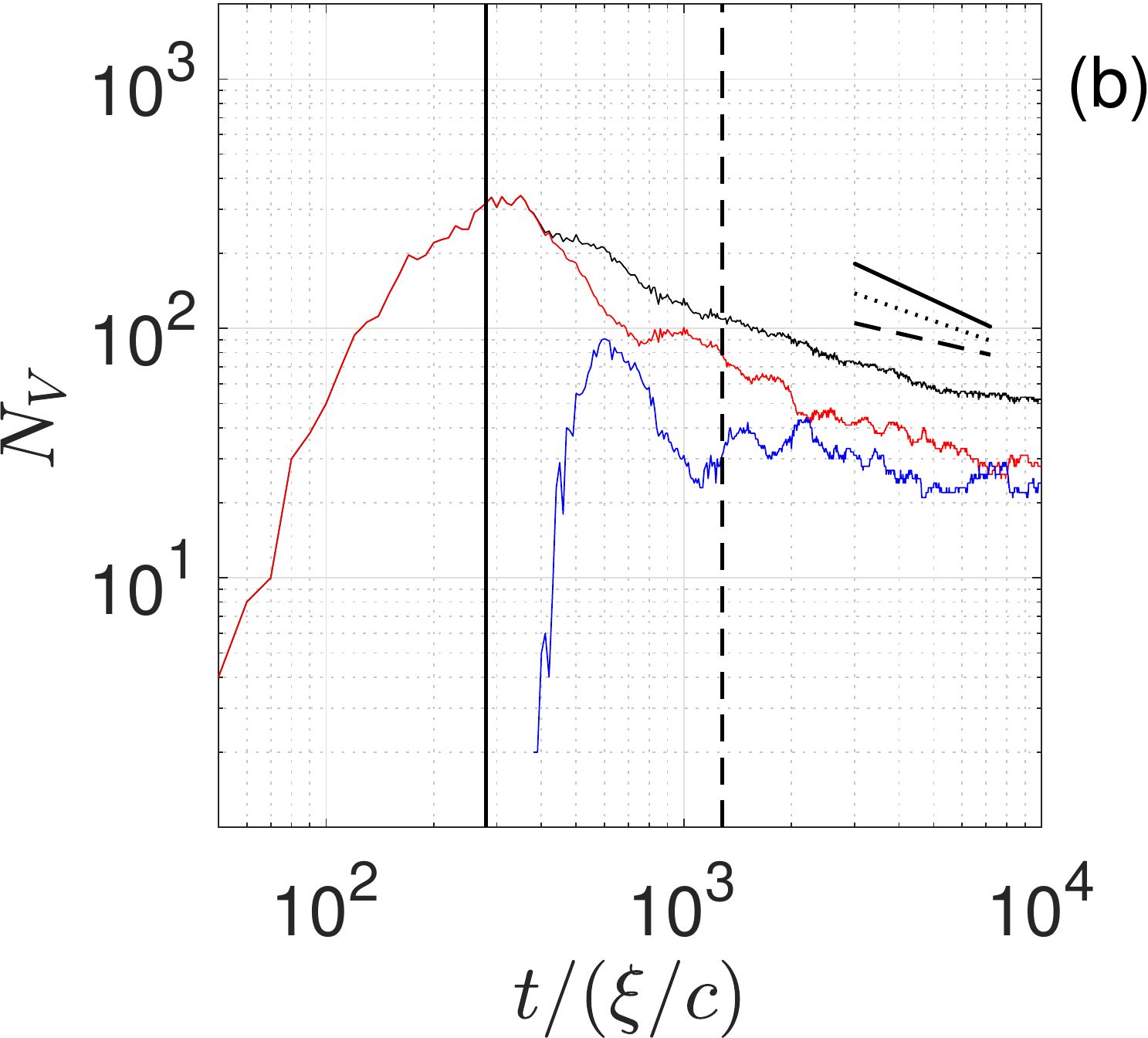}
\includegraphics[width=0.32\linewidth]{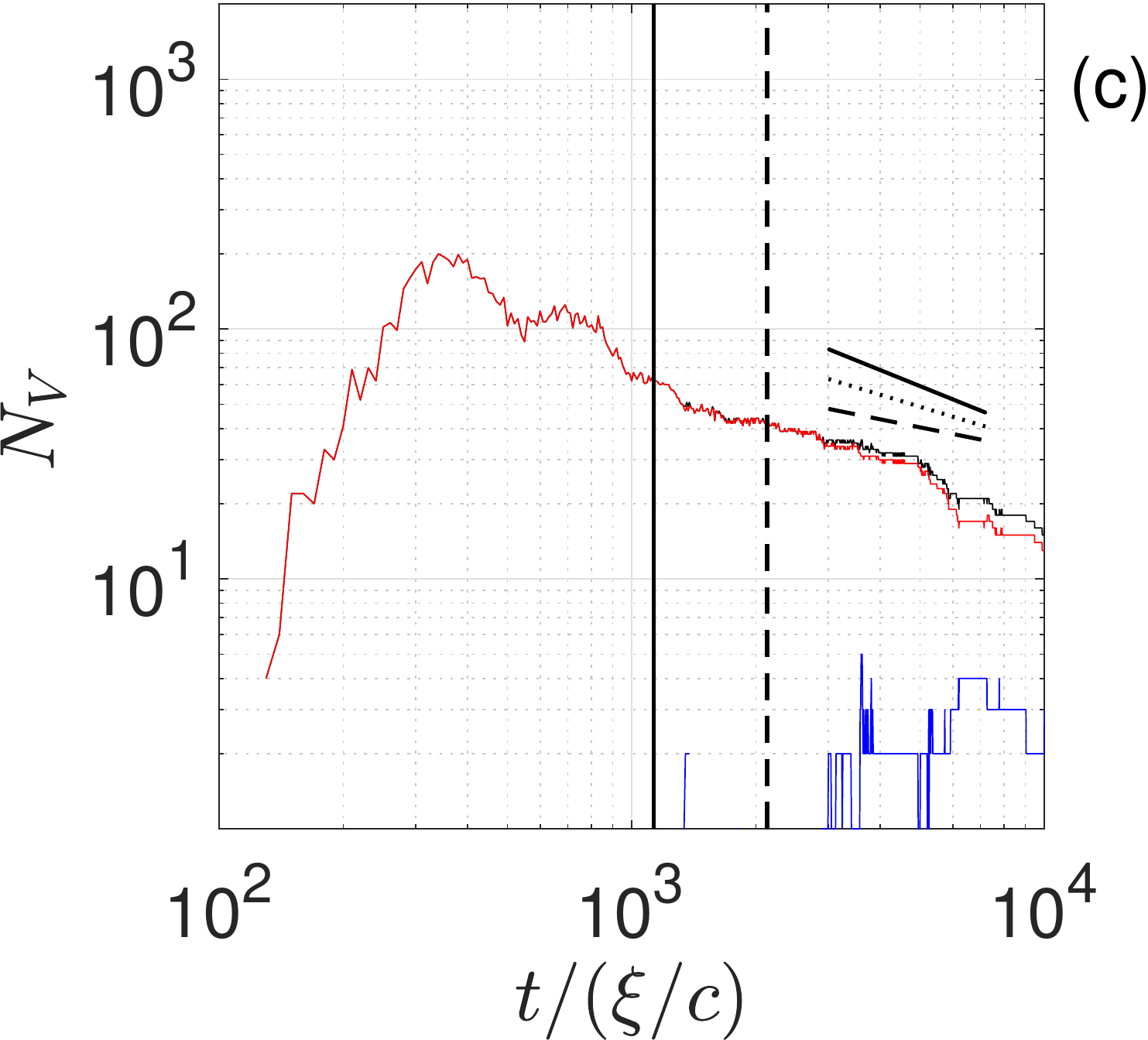}
\includegraphics[width=0.32\linewidth]{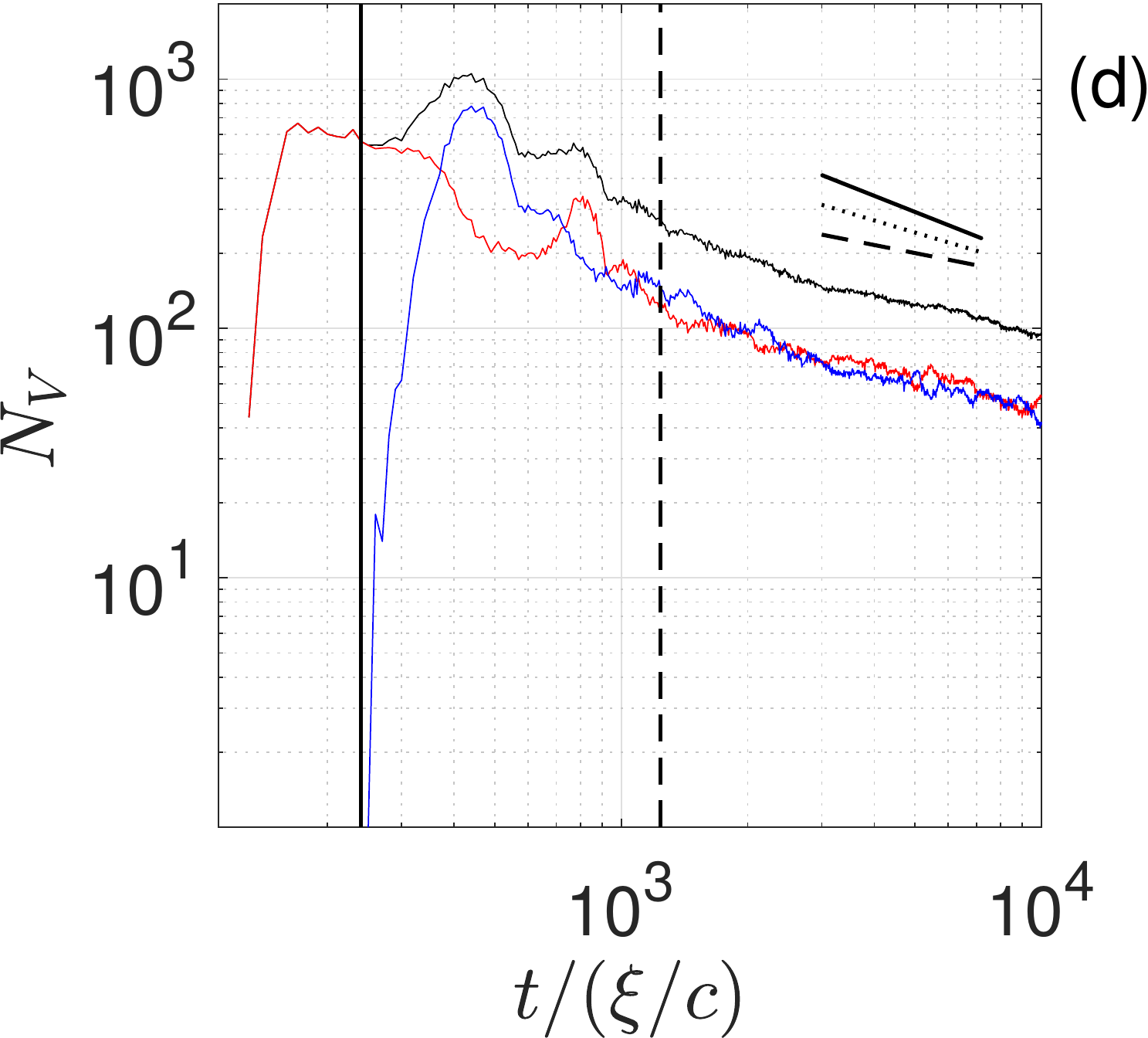}
\includegraphics[width=0.32\linewidth]{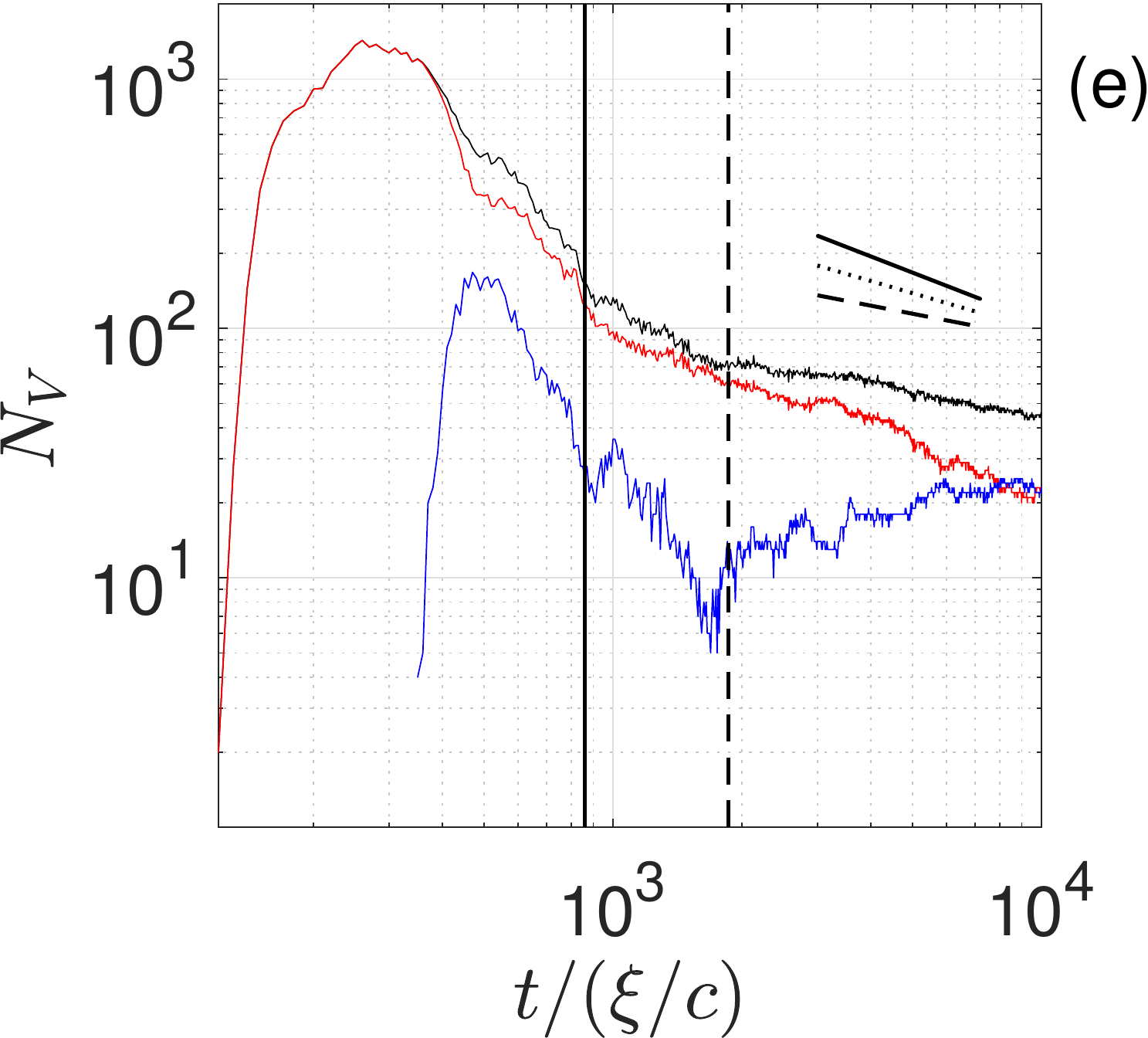}
\includegraphics[width=0.32\linewidth]{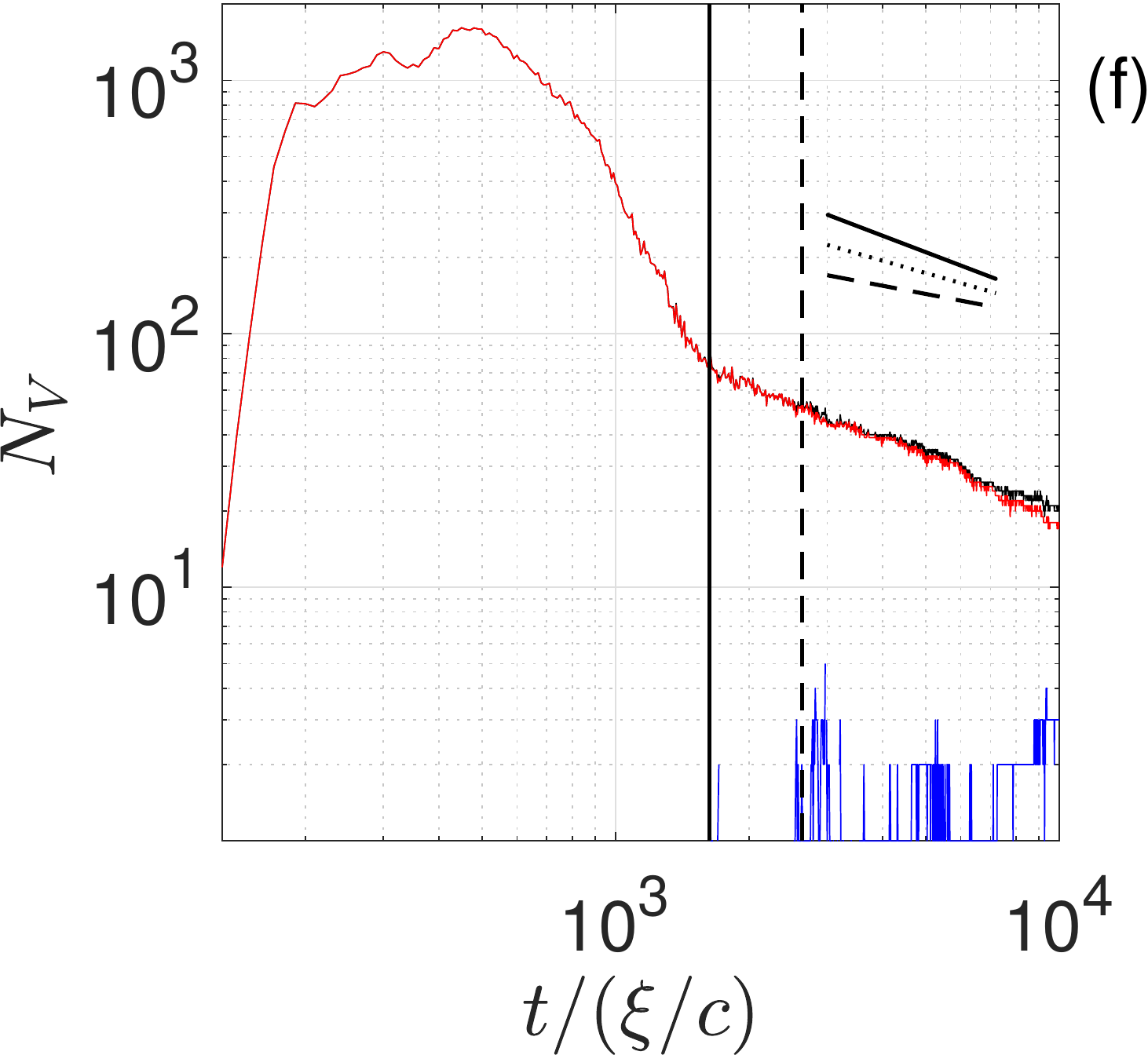}
 \caption{{\it (colour online)} Log-log plots of number of vortices against time. 
 For low initial imbalance, (a-c), $Z_0 =0.49$ and barrier height: (a) $V_0=0.0$; (b) $V_0=0.6$ and (c) $V_0=1.5$. For high initial imbalance, (c-g), $Z_0 =0.88$ and barrier height: (c) $V_0=0.0$; (e) $V_0=0.6$ and (f) $V_0=1.5$. The black full line is  shows $t^{-2/3}$ law, the dotted to $t^{-1/2}$ and the dashed line -- $t^{-1/3}$. The vertical black line shows the time when $Z(t)<0.05$ for the first time. The red lines corresponds to the number of vortices in the right box, the blue the left box and the black the total number of vortices.}\label{NVG}
\end{figure}


\begin{align}
\frac{d N_v}{dt}=-CN_v^{\alpha},\label{log}
\end{align}
where $\alpha$ here corresponds to the number of colliding vortices causing an annihilation and C is a constant. Equation \eqref{log} is a crude approximation which does not take into account any correlation between the vortices, nor does it take into account spatial inhomogeneity in the system. Following this approach, for two-vortex annihilations we have $N_v$ is proportional to $t^{-1}$, for three-vortex interactions -- to $t^{-1/2}$ and four-vortex interactions -- to $t^{-1/3}$.

\begin{figure}[ht]
\centering
\includegraphics[width=0.3\linewidth]{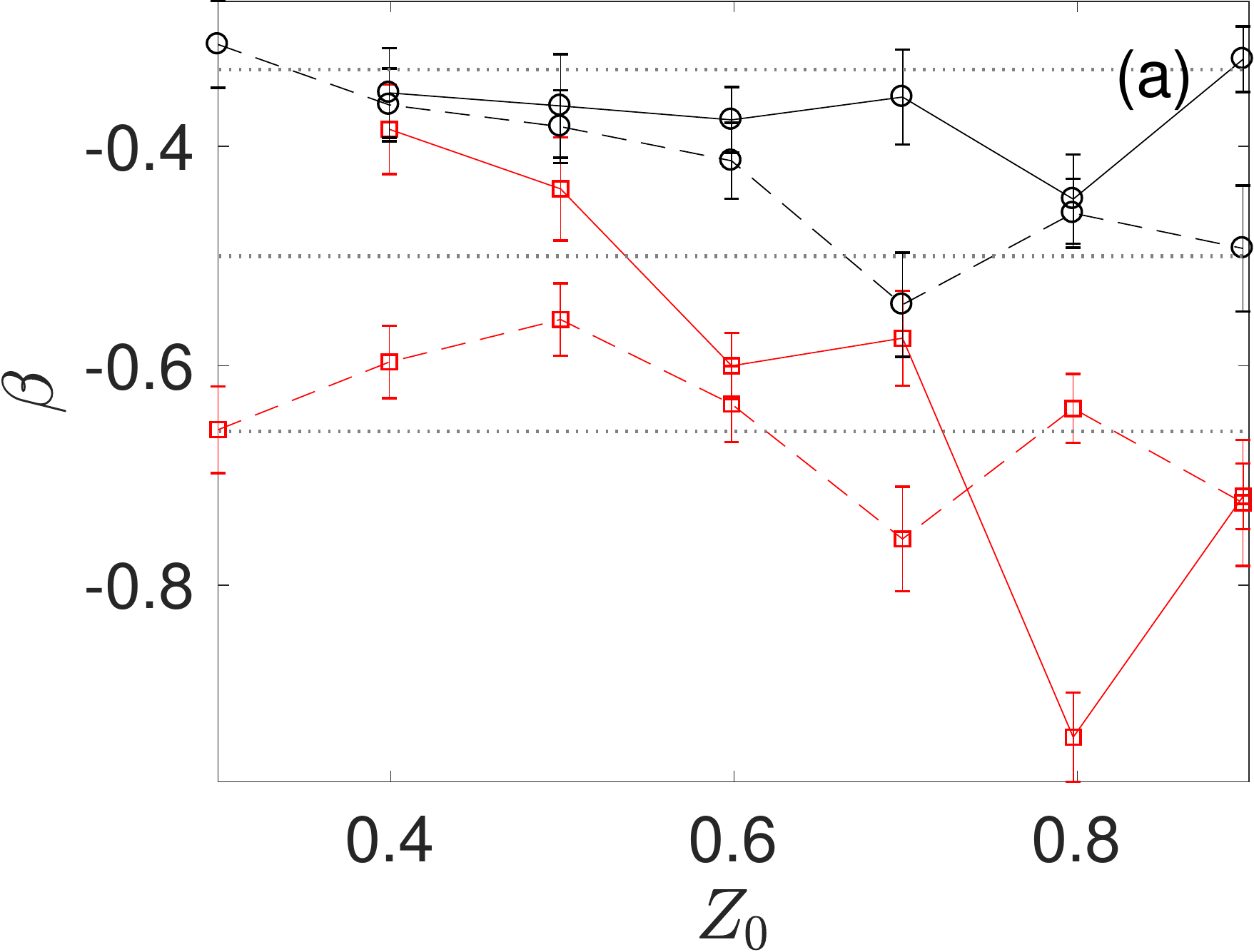}
\includegraphics[width=0.3\linewidth]{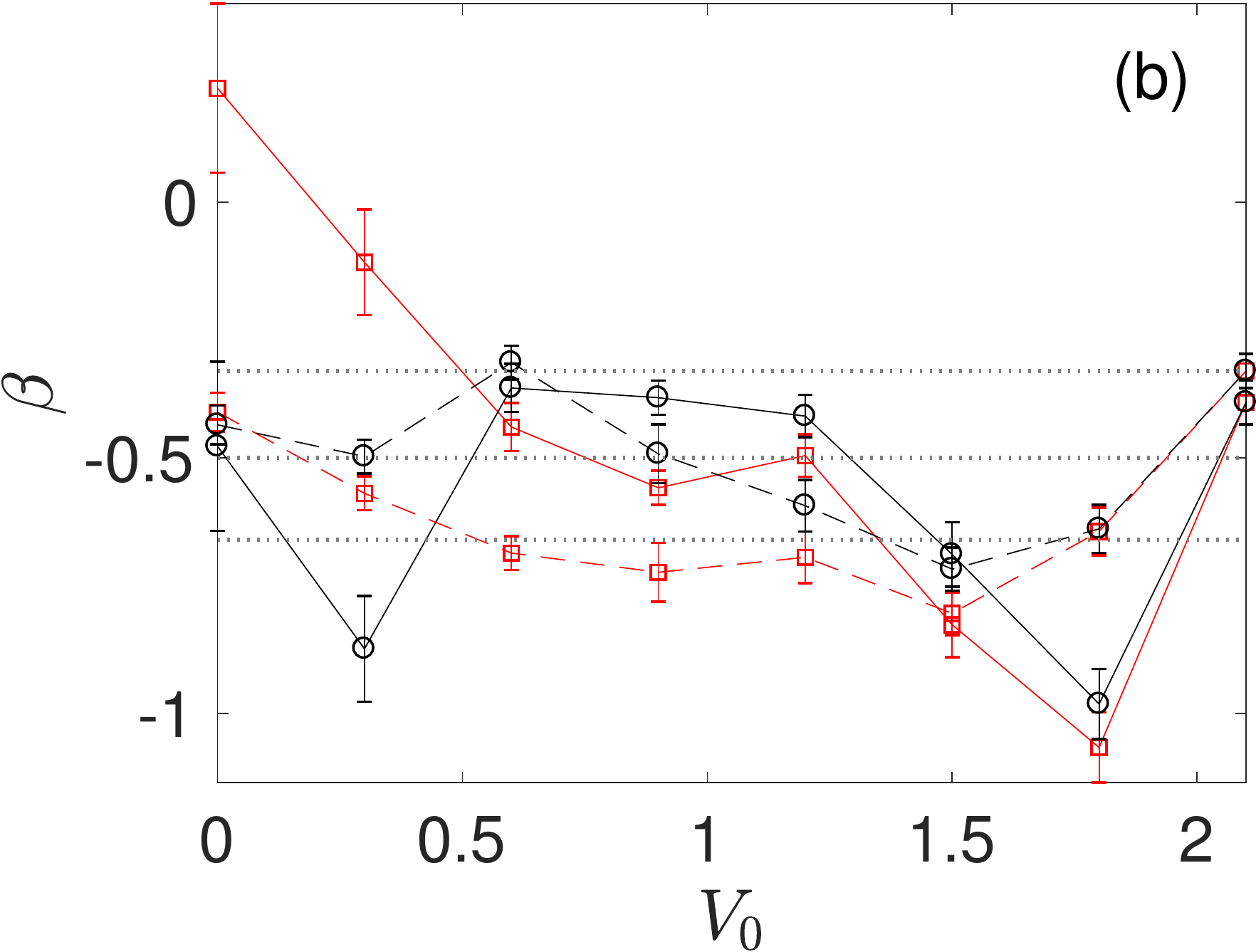}
\caption{{\it (colour online)} Vortex number decay fitted with $t^{-\beta}$. Panel (a): The decay rate $\beta$ in the entire domain (black circles) and right box (red squares) as we increase initial imbalance with $V_0=0.6/\mu$ (full line) $0.9/\mu$ (dashed line). Panel (b): The decay rate $\beta$ in the entire domain (black circles) and right box (red squares) as we increase barrier height with initial imbalance $Z_0=0.49$ (full line) $0.88$ (dashed line). The grey dotted horizontal lines are markers for $-1/3$, $-1/2$ and $-2/3$.}\label{NVM}
\end{figure}

In our system, the decay of vortices from $B_R$ can happen in four ways: (1) vortices annihilate via vortex-vortex interaction; (2) vortices annihilate at the barrier or boundary, see Fig.~\ref{DNO} (c); (3) vortices pass the barrier and enter to $B_L$, see Fig.~\ref{DNO} (a,b,d) (however, the barrier can be chosen such that the vortices cannot penetrate or so that only vortex dipoles of a certain size can exist in the left well, this is discussed further in section \ref{VDSWB}); (4) vortices annihilate from interacting with the background sound.

\begin{figure}[ht]
\centering
\includegraphics[width=0.45\linewidth]{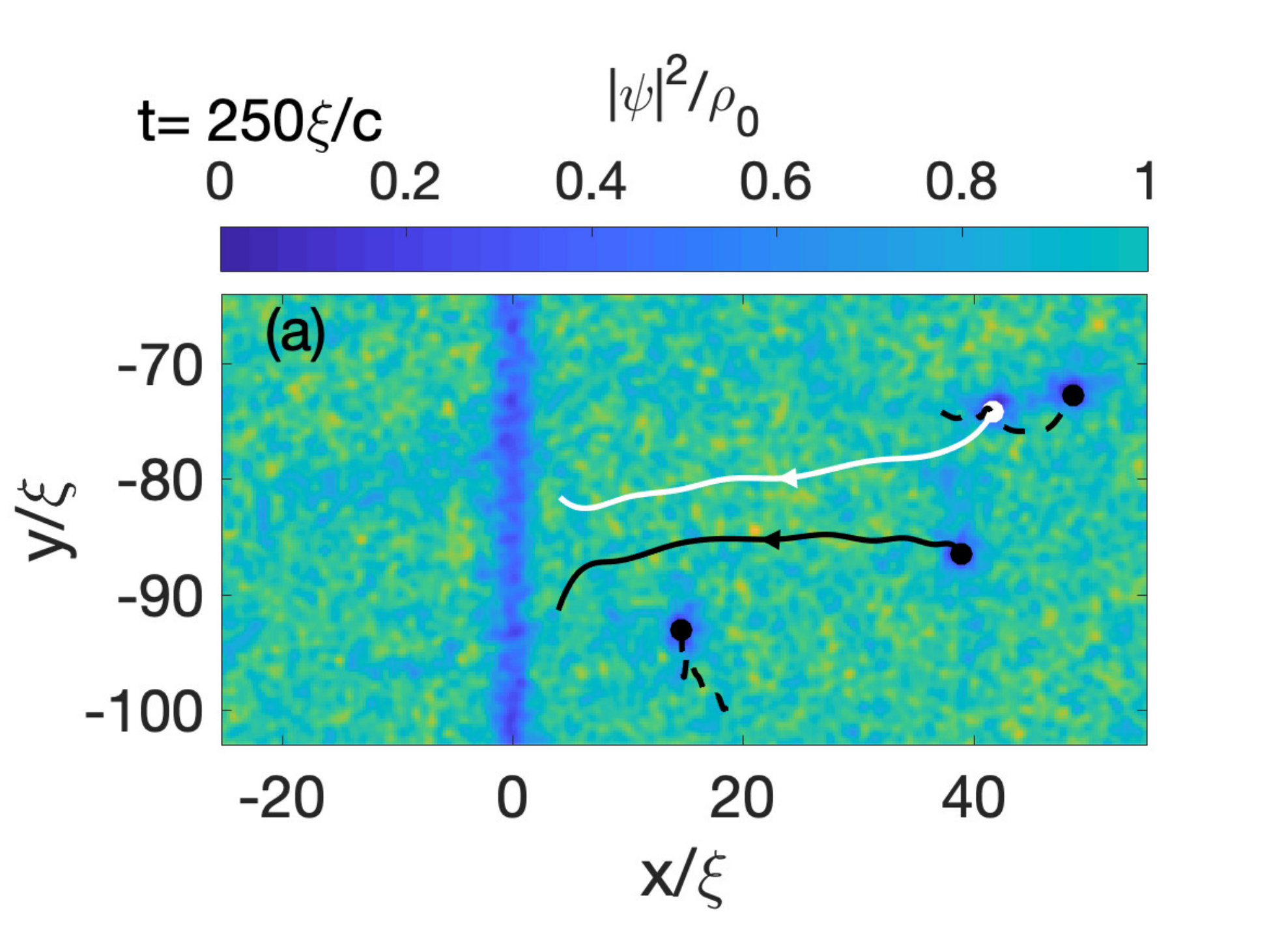}
\includegraphics[width=0.45\linewidth]{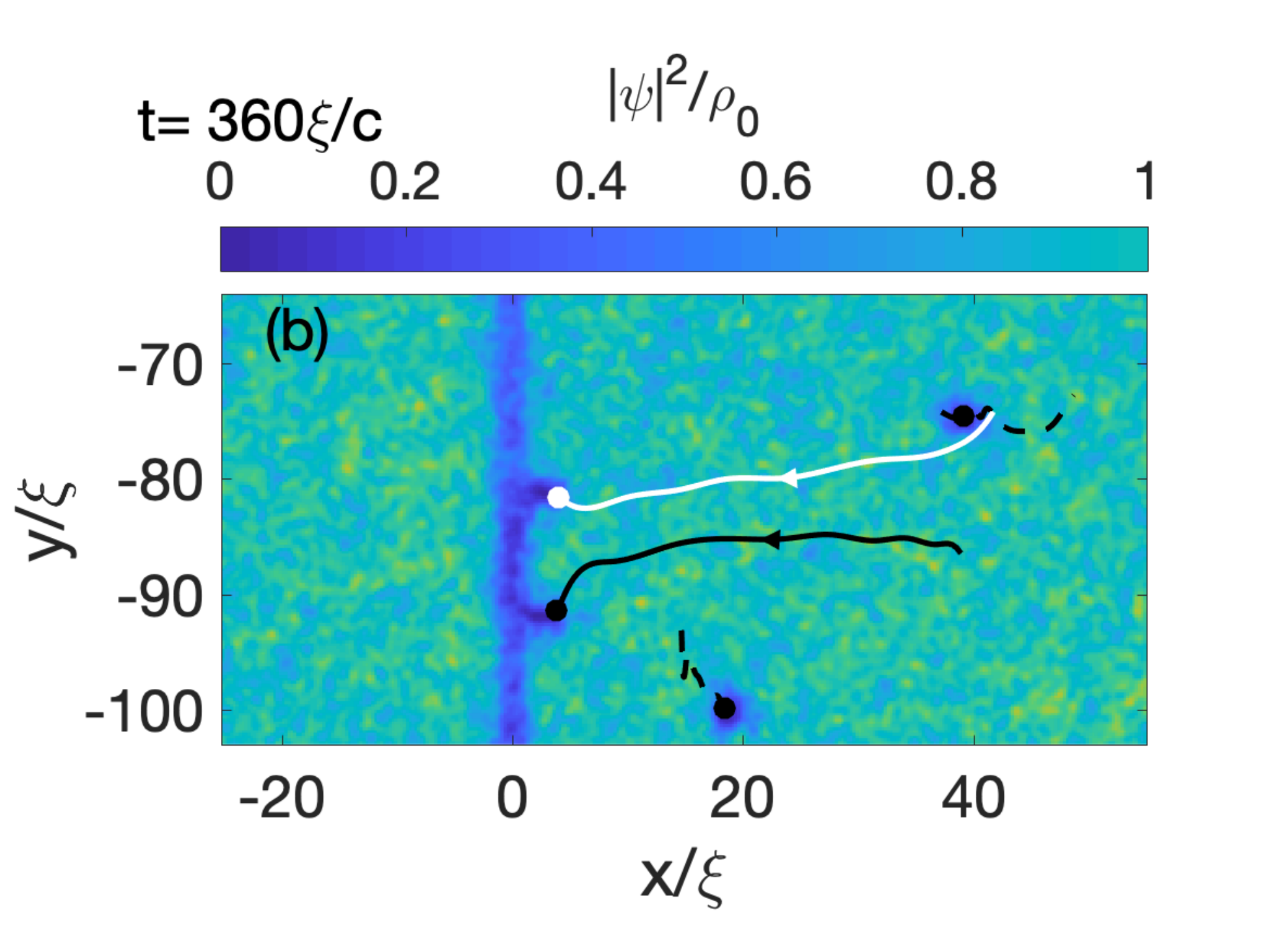}
\includegraphics[width=0.45\linewidth]{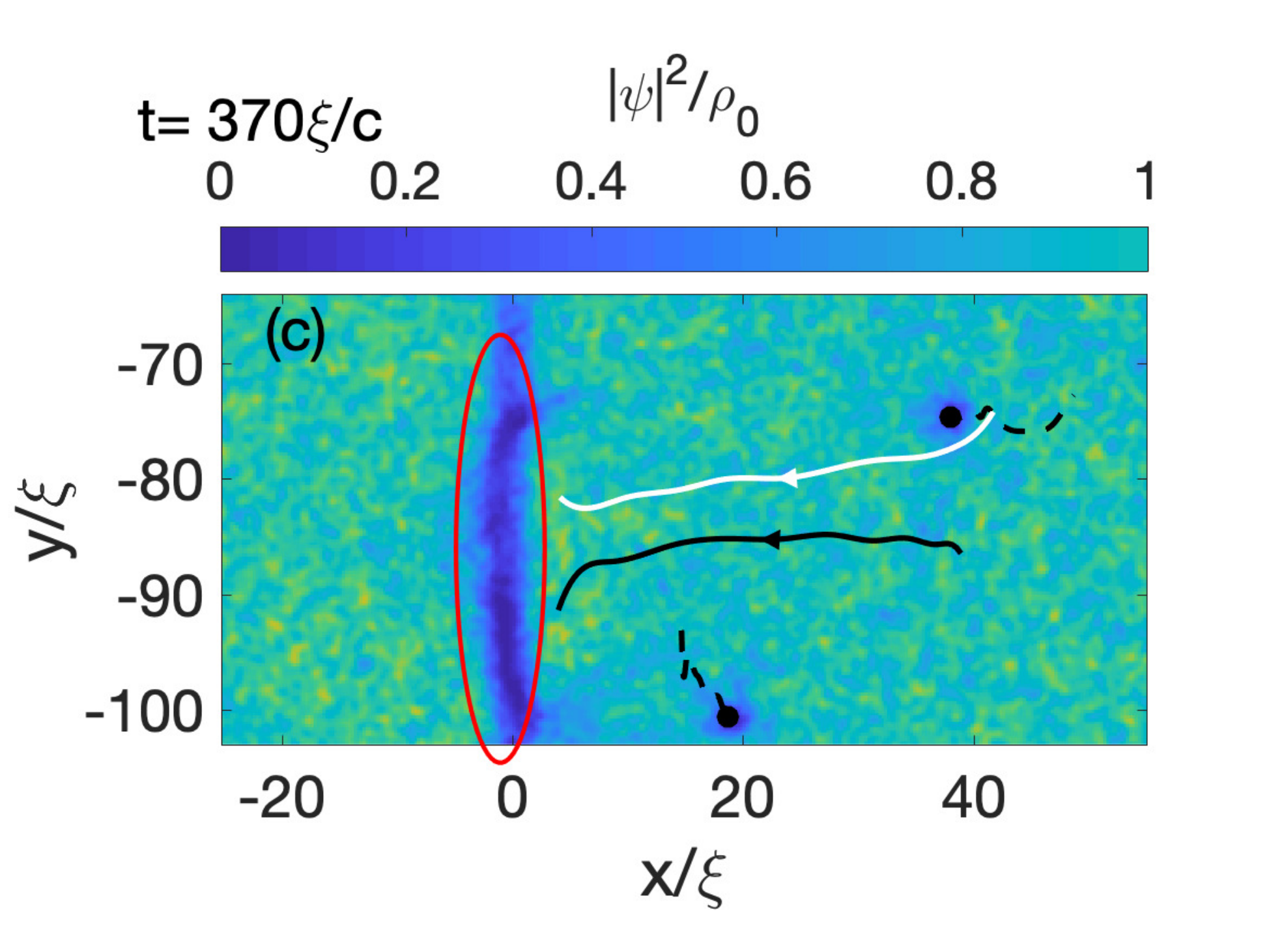}
\includegraphics[width=0.45\linewidth]{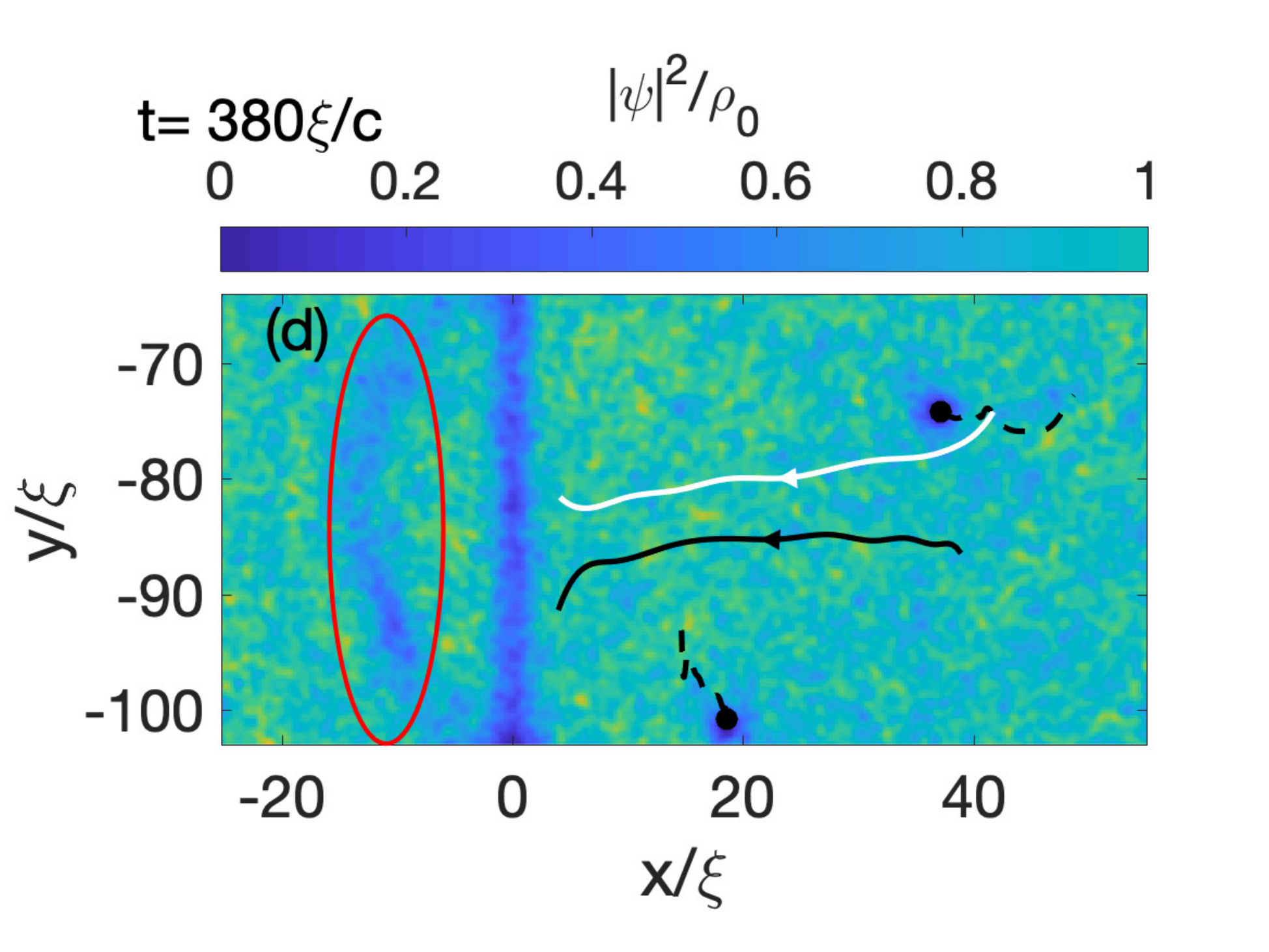}
\caption{{\it (colour online)} Vortices annihilating at the barrier. Density plots with parameters $V_0=1.2/\mu$, $\sigma=1.2/\xi$ and $Z_0 = 0.49/\mu$. The time is indicated in the figures. The white line is the trajectory of the vortex and the black lines are the  antivortex trajectories for the vortices which collide with the barrier. The filled circles are the positions of the vortices in each frame. The arrows indicate the direction of motion of the vortices and the ellipses the sound waves post-collision. The dashed trajectories are of the vortices which do not interact with the barrier.}\label{ANN}
 \end{figure}

Fig.~\ref{NVG} shows examples of the evolution of the number of vortices in a log-log plot. In the figure the red line represents the number of vortices in the right box $N_{V_R}$, the blue the number in the left box $N_{V_L}$ and the black the total in the entire domain $N_{V_R}+N_{V_L}$. We overlay a vertical line for which the imbalance has become almost zero ($Z<0.05$) for the first time in each simulation, this is an indicator of when the initial vortex creation period has ended.


We first focus on the decay rate in the later stages of the dynamics where we expect to see the exponents predicted above. We present in Fig.~\ref{NVG} cases with two values of initial imbalance, $Z_0=0.49$ (a-c) and $Z_0=0.88$ (d-f). In the cases with no barrier (a, e) we see that the vortices move freely between the two boxes. In Fig.~\ref{NVG} (a) in particular, we do not produce enough vortices to see vortex turbulence, thus we do not see a clear decay rate. In Fig.~\ref{NVG} (e) homogeneous vortex turbulence develops quickly with an almost equal amount of vortices in each box. In this case we see a decay closer to $t^{-1/3}$ which is predicted for a four-vortex process. In Fig.~\ref{NVG} (f), for small value of $V_0$,  see also a decay closer to $t^{-1/3}$, whereas a  $t^{-1}$ decay is seen  for higher barrier height in Fig.~\ref{NVG} (g), which corresponds to a two-vortex collision in the logistic equation.

Clearly the amount of vortices in the left box decreases with the barrier height. It also appears that the number of vortices in the right box remains almost the same whilst the number in the left tends to zero. We also note that for cases with a high initial amount of vortices there is a steeper decay rate. We propose that some transition in the dominant type of the vortex collision occurs at a vortex density which is dependent on the mean density  of the vortices. Indeed, it is natural to think that the higher-order vortex collision (e.g. four-vortex) dominates over the lower-order collision (e.g. three-vortex) for higher vortex densities and vice versa.  The mean vortex density is dependent on the barrier height in a non-trivial way as shown in section \ref{VDSWB}.

It is difficult to distinguish the best fit from the decay in Fig.~\ref{NVG} by visual inspection. We instead calculate the best fit numerically and present the results in Fig.~\ref{NVM}. To approximate the decay rate from our simulated data we also need to approximate the time at which the vortex production ends. To ensure that we have safely passed the initial vortex proliferation, we choose that the time interval begins when the density imbalance $Z(t)$ first falls below zero plus some additional buffer time ($1000\xi/c$) and continues until the final time $t = 10000\xi/c$. We present the beginning of the time interval as the vertical dashed line in Fig.~\ref{NVG}. We then calculate a least squares linear fit, the gradient of which is the approximated decay rate over the chosen time interval. Along with the decay rate we present the average error of the fits from the data in the form of error bars.

In Fig.~\ref{NVM} (a) we present the exponent plotted against increasing initial imbalance $Z_0$. We notice that the decay rate is faster for a higher imbalance. We conjecture that this is due to the higher mean vortex density as well as an increased acoustic component (see Fig.~\ref{NV} (c)) which interacts with the vortices helping them to annihilate. In Fig.~\ref{NVG} (b) as the barrier height increases the total number of vortices (black line) and the vortices in the right box (red line) converge due to all of the vortices being in the right box. We also see in Fig.~\ref{NVM} (a) and (b) that the decay rate fluctuates. However, for most of the parameters, it seems to be steeper than $t^{-1/2}$ and shallower than $t^{-1}$ which may be due to the combination of all of the annihilation mechanisms. For instance, the decay rate may be explained by the interaction with the boundary and barrier, which can aid annihilation.
For instance in Fig.~\ref{ANN} we see a vortex dipole scatter of a third vortex  (a three-vortex precess) until the dipole is small enough such that the boundary or the barrier will annihilate them (a two-vortex process). Also in Fig.~\ref{ANN} (c,d) the red ellipses show the dynamics during (c) and after (d) the interaction with the barrier. We highlight that the rarefaction pulse caused by the annihilation is directed into the left box, thus the pulse is not likely to interact with other vortices causing more annihilations. We note that by increasing dipole size, we do not necessarily increase the chance of annihilation; see Fig.~\ref{DVV} (a). If we take $V_0=1.6/\mu$ we see that smaller dipoles are trapped whereas larger dipoles annihilate. 

  

\section{Conclusions}

In this paper, we have explored the use of a Josephson junction set-up for generating BEC vortex turbulence.
We have shown that the generation and decay of vortices in a Josephson junction BEC configuration can be altered by controlling the barrier height and the initial density imbalance parameters, which can be readily applied to existing experimental apparatus. We discussed the critical advantage of this method for generating vortex turbulence is the creation of vortices in low-density regions, with the density then being increased to solidify (shrink) the vortex cores. We have provided ranges of parameters to tune the barrier to produce a certain optimal number of vortices and also shown parameters which allow for vortex penetration. We showed that for a higher imbalance we produce more vortices. However, most of these vortices decay quickly leaving much acoustic noise and large-scale density sloshing, and for this reason it may be preferential to have a smaller initial imbalance. For instance, if one would like to create vortices confined to a single box with the acoustic component less than 10\% of the total energy one would choose an initial imbalance of $Z_0=0.6$, with a barrier with strength (measured in chemical potential $\mu$) in the range of $\mu$ to $ 1.5\mu$ with the width of the order of the healing length. Although our simulations are larger than current experiments, there are experiments that are not so much smaller \cite{Gauthier:2019aa,Johnstone:2019aa}.  \\
 
Another disadvantage of the vortex turbulence without the barrier is the large wave-vortex interaction. A key consideration is the control of the amount of sound energy released as part of the snake instability. Such sound can be shared over the two boxes while confining the vortices to one box, and thereby separate the vortices and sound in physical space better so that the sound does not adversely affect the vortex dynamics. We show that the interaction of vortices with a quasi-1D barrier is non-trivial and that the barrier also can work as another effective mechanism for vortex dissipation. This is particularly interesting in terms of using such a barrier to filter vortex turbulence and produce vortex dipoles of a certain size. For instance, one could design a barrier, or a set of barriers, such that only dipoles with separations greater (as smaller ones  annihilate on the barrier) than $d_{min}$ and smaller (as larger ones are trapped or annihilate) than $d_{max}$ are produced; however, this would require a more detailed study which will be the subject of future work.  In this work we focused on the vortex turbulence; however, the acoustic turbulence present is also worth further study. 


\begin{acknowledgments}
AG is supported by a EPSRC studentship Ref~1642416. AG, and DP were supported by the EPSRC First Grant (ref. EP/P023770/1). SN is supported by the Chaire D'Excellence IDEX (Initiative of Excellence) awarded by Universit\'e de la C\^ote d’Azur, France.
This project has received funding from the European Union’s Horizon 2020 research and innovation programme under the grant agreements No 823937 in the framework of Marie Skłodowska-Curie HALT project and No 820392 in the FET Flagships PhoQuS project.
 Computations were carried out on the High Performance Computing Cluster and supported by the Scientific Computing Research Technology Platform at the University of Warwick. 
\end{acknowledgments}

\bibliographystyle{phjcp}
\bibliography{references}

\end{document}